\documentclass[apj]{emulateapj}

\usepackage{amsmath}
\usepackage{graphicx}
\usepackage{amssymb}
\usepackage{bm}
\usepackage{hyperref}
\usepackage{epstopdf}
\usepackage{color}

\newcommand{\msunh}{\>h^{-1}\rm M_\odot}
\newcommand{\Lsunhh}{\,h^{-2}\rm L_\odot}

\newcommand{\mpch}{\>h^{-1}{\rm {Mpc}}}
\newcommand{\kms}{\>{\rm km}\,{\rm s}^{-1}}
\newcommand{\kmsmpc}{\>{\rm km}\,{\rm s}^{-1}\,{\rm Mpc}^{-1}}

\newcommand{\rmd}{{\rm d}}

\usepackage{color}

\shorttitle{Galaxy groups/clusters in the DESI Legacy Imaging Surveys}
\shortauthors{Yang et al.}

\begin{document}
            

\title{An Extended Halo-based Group/Cluster finder: application to the DESI legacy imaging surveys DR8}
    
\author{Xiaohu Yang\altaffilmark{1,2}, 
  Haojie Xu\altaffilmark{1}, Min He\altaffilmark{1}, Yizhou Gu\altaffilmark{1}, Antonios Katsianis\altaffilmark{2}, Jiacheng Meng\altaffilmark{3}, Feng Shi\altaffilmark{4},  Hu Zou\altaffilmark{5}, 
  Youcai Zhang\altaffilmark{6}, Chengze Liu\altaffilmark{1}, Zhaoyu Wang\altaffilmark{1}, Fuyu Dong\altaffilmark{7}, Yi Lu\altaffilmark{6}, Qingyang Li\altaffilmark{1}, Yangyao Chen\altaffilmark{3,10}, Huiyuan Wang\altaffilmark{8,9}, Houjun Mo\altaffilmark{10}, Jian Fu\altaffilmark{6},  Hong Guo\altaffilmark{6}, Alexie Leauthaud\altaffilmark{11}, Yu Luo\altaffilmark{12}, Jun Zhang\altaffilmark{1}, Ying Zu\altaffilmark{1} }

\altaffiltext{1}{Department of Astronomy, School of Physics and
  Astronomy, and Shanghai Key Laboratory
  for Particle Physics and Cosmology, Shanghai Jiao Tong University, Shanghai 200240, China;
  E-mail: xyang@sjtu.edu.cn}

\altaffiltext{2}{Tsung-Dao Lee Institute, and Key Laboratory for
    Particle Physics, Astrophysics and Cosmology, Ministry of Education, Shanghai Jiao Tong University,
  Shanghai 200240, China}
  
\altaffiltext{3}{Department of Astronomy, Tsinghua University, Beijing 100084, China}  

\altaffiltext{4}{Korea Astronomy and Space Science Institute, Yuseong-gu, Daedeok-daero 776, Daejeon 34055, Korea} 

\altaffiltext{5}{Key Laboratory of Optical Astronomy, National Astronomical Observatories, Chinese Academy of Sciences, Beijing 100012, China}  

\altaffiltext{6}{Key Laboratory for Research in Galaxies and Cosmology,
  Shanghai Astronomical Observatory; Nandan Road 80, Shanghai 200030,
  China}  

\altaffiltext{7}{School of Physics, Korea Institute for Advanced Study, 85 Heogiro, Dongdaemun-gu, Seoul, 02455, Republic of Korea}

\altaffiltext{8}{Key Laboratory for Research in Galaxies and Cosmology, Department of Astronomy, University of Science and Technology of China, Hefei, Anhui 230026,  China}

\altaffiltext{9}{School of Astronomy and Space Science, University of Science and Technology of China, Hefei Anhui 230026, China}

\altaffiltext{10}{Department of Astronomy, University of Massachusetts Amherst, MA 01003, USA}    

\altaffiltext{11}{Department of Astronomy and Astrophysics, University of California, Santa Cruz, 1156 High Street, Santa Cruz, CA 95064, USA}  

\altaffiltext{12}{Purple Mountain Observatory, No. 10 Yuanhua Road, Nanjing 210033, China}


\begin{abstract}
We extend the halo-based group finder developed by \citet[][]{Yang2005a} to use data {\it simultaneously} with either photometric or 
 spectroscopic redshifts. A mock galaxy redshift survey constructed from 
 a high-resolution N-body simulation is used to evaluate the performance of this extended group finder.  For galaxies with magnitude ${\rm z\le 21}$ and redshift $0<z\le 1.0$ in the DESI legacy imaging surveys (the Legacy Surveys), our group finder successfully identifies more than 60\% of the members in about $90\%$ of halos with mass  $\ga 10^{12.5}\msunh$. Detected groups with mass $\ga 10^{12.0}\msunh$ have a purity (the fraction of true groups) greater than 90\%. The halo mass assigned to each group has an uncertainty of about 0.2 dex at the high mass end $\ga 10^{13.5}\msunh$ and 0.40 dex at the low mass end. Groups with more than 10 members have a redshift accuracy of $\sim 0.008$. We apply this group finder to the Legacy Surveys DR8 and find 5.2 Million groups with at least 3 members. About 387,000 of these groups have at least 10 members. The resulting catalog containing 3D coordinates, richness, halo masses, and total group luminosities, is made publicly available.  
\end{abstract}




\section{Introduction}

The past two decades have seen great progress in establishing the connection between galaxies and dark matter halos, as parameterized via
either the conditional luminosity function (CLF) or the halo occupation distribution  \citep[HOD; e.g.,][]{Jing1998, 
 Peacock2000, Yang2003, vandenBosch2003, vandenBosch2007, Zheng2005, Tinker2005,  Mandelbaum2006, Brown2008, More2009, Cacciato2009, 
Neistein2011, Avila-Reese2011, Leauthaud2012}. The galaxy-dark matter connection describes how galaxies with different properties occupy halos of different mass and contains important information about how galaxies form and evolve in dark matter halos \citep[see][for a concise
review]{Mo2010}.   

Apart from establishing these empirical models, a more direct way of studying the galaxy-halo connection is using galaxy groups, provided that these galaxy groups are defined as sets of galaxies that reside in the same dark matter halos.\footnote{In this paper, we refer to a system of galaxies as a group regardless of its richness, including isolated galaxies (i.e., groups with a single member) and rich clusters of galaxies.}  Using a well-defined galaxy group catalog, we can not only study the properties of galaxies as a function of their group properties or how the member galaxies evolve within different environments \citep[e.g.,][]{Yang2005c, Collister2005, vandenBosch2005, Robotham2006, Zandivarez2006, Weinmann2006} but also probe how dark matter halos trace the large-scale structure of the universe \citep[e.g.][]{Yang2005b, Yang2006, Coil2006}.

Thanks to many large scale surveys that have been carried out during the past few decades, numerous group catalogs have been
constructed, such as the CfA redshift survey \citep[e.g.,][]{Geller1983}, the  Las Campanas Redshift Survey  \citep[e.g.,][]{Tucker2000}, the  2-degree Field Galaxy Redshift Survey  \citep[e.g.,][]{Merchan2002, Eke2004, Yang2005a,  Tago2006, Einasto2007}, the high-redshift DEEP2 survey \citep[][]{Gerke2005}, the Two Micron All Sky Redshift Survey  \citep[2MASS; e.g.,][]{Crook2007, Diaz-Gimenez2015, Lu2016, Lim2017}, the zCOSMOS \citep{Wang2020} and most notably the Sloan  Digital Sky Survey  (SDSS). Based on SDSS observations, various group catalogs have been constructed with a friends-of-friends  (FOF) algorithm \citep[e.g.,][]{Goto2005, Berlind2006, Merchan2005}, the C4 algorithm \citep[e.g.,][]{Miller2005}, and the halo-based group finder developed in \citet{Yang2005a}  \citep[e.g.,][]{Weinmann2006, Yang2007, Yang2012, Duarte2015, Rodriguez2020}. Among these group finders, the halo-based group finder established in \citet{Yang2005a, Yang2007} has the particular advantage that it links galaxies to their common dark matter halos \citep[e.g.,][]{Campbell2015}. Along this line, efforts were also made to improve the halo mass estimations \citep[e.g.,][]{Lu2015, Wang2020, Tinker2020}.

Based on the galaxy groups constructed from large redshift surveys, nowadays we can directly measure the halo occupation distribution or the  conditional luminosity functions  of galaxies in halos of different masses \citep[e.g.][]{Yang2005c, Yang2008, Yang2009, Rodriguez2015, Lan2016}, study the dependence of galaxy properties on their host halos, e.g. the {\it galactic conformity} found 
by \citet{Weinmann2006} \citep[see also][]{Knobel2015, Kawinwanichakij2016, Darvish2017} or the halo and stellar mass quenching factors disentangled by \citet{Wang2018}, and measure the group-galaxy cross-correlation function to evaluate how galaxies are distributed within and beyond their host halos \citep[e.g.][]{Yang2005d, Coil2006, Knobel2012}. Since galaxy groups trace the dark matter halos we can also stack groups with similar masses, to probe the weak signals of Sunyaev-Zel'dovich (SZ) effects  \citep[e.g.][]{Li2011, Vikram2017, Lim2018, Lim2020} and weak gravitational lensing signals \citep[e.g.][]{Mandelbaum2006, Yang2006, Han2015, Viola2015, Luo2018} over a large halo mass range.

In addition to the large spectroscopic redshift surveys, which are usually expensive and 
observationally time-consuming, photometric
redshift surveys in recent years have provided two orders of magnitude more galaxies for our studies. Group/cluster catalogs have also been constructed  from  these photometric redshift data, especially from SDSS observations. For example, \citet{Goto2002} developed  a  
cut-and-enhance method and applied  it to the early SDSS  commissioning data.  \citet{Bahcall2003} used two different selection methods including a hybrid matched filter method \citep{Kim2002}  and a ``maxBCG'' method to find clusters.  \citet{Lee2004}  identified compact 
groups in the SDSS Early  Data  Release. \citet{Koester2007} identified clusters using the maxBCG red-sequence method from the SDSS DR5. citet{Liu2008} developed a probability FOF method 
and tested it using DEEP2 mock catalogs.
\citet{Hao2010} used a GMBCG method to find clusters in SDSS DR7. \citet{Szabo2011} constructed a cluster catalog from SDSS DR6 using an 
adaptive matched filter cluster finder. \citet{Wen2012} created a cluster catalog using the photometric redshifts of galaxies from SDSS III. \citet{Rykoff2014} constructed a cluster catalog using the red-sequence cluster finder, redMaPPer, from SDSS DR8. 
\citet{Oguri2018} constructed a  cluster catalog from the Hyper Suprime-Cam (HSC) Subaru Strategic Program  using the CAMIRA algorithm. 

The above group/cluster-finding algorithms applied to the photometric redshift data are mainly based on either the red-sequence feature or the overdensity feature of the clusters, which require either the existence of the brightest cluster galaxy or an overdensity peak as the candidate cluster location. Thus, the group/cluster catalogs constructed may suffer from incompleteness due to the lack of BCG observation, or the threshold of the overdensity, or contamination caused by the foreground and background galaxies. In general, only
the rich clusters of these catalogs are reliable tracers of the overall cluster population. In this work, to make better use of the available photometric redshift survey data, we propose to extend the halo-based group finder developed in \citet{Yang2005a, Yang2007} (hereafter, Y05 and Y07), so that it can be applied to both photometric and spectroscopic redshift surveys. We will see that this extended group/cluster finder has  
better performance for the group completeness which is important for cosmological studies and better performance for the group membership determination which is important for galaxy formation studies. After testing our extended group finder on the mock photometric redshift galaxy survey, we employed it to the currently largest photometric redshift galaxy sample, the DESI legacy image surveys (hereafter the Lecacy Surveys)
DR8 galaxy catalog, and generated group/cluster catalogs for subsequent investigations.

This paper is organized as follows: In  Section \ref{sec_data} we provide a detailed description of the data we use for our investigation, 
including the Legacy Surveys DR8 galaxy catalog and the mock redshift surveys. In Section~\ref{sec_steps} we describe the main steps adopted for our extended halo-based group finder.  The performance of this group finder is tested in Section \ref{sec_test}. In Section ~\ref{sec_catalogue} we present some of the basic properties of the groups we extract from the Legacy Surveys DR8. Finally, we summarize our results in Section~\ref{sec_conclusion}.  For the observational data, we adopt a $\Lambda$CDM cosmology with parameters that are consistent with the Planck 2018 results \citep[][hereafter Planck18 cosmology]{Planck2018}: $\Omega_{\rm m} = 0.315$, $\Omega_{\Lambda} = 0.685$, $n_{\rm s}=0.965$, $h=H_0/(100 \kmsmpc) = 0.674$ and $\sigma_8 = 0.811$.

\section[]{The dataset}
\label{sec_data}

In this section, we describe the data used in this study: (i) the galaxy catalog with photometric redshifts obtained from the Legacy Surveys and the spectroscopic redshifts from various sources, (ii) the mock galaxy catalogs generated from N-body simulations to test the performance of our group/cluster finder.

\begin{figure*}
\center
\includegraphics[height=7.0cm,width=14cm,angle=0]{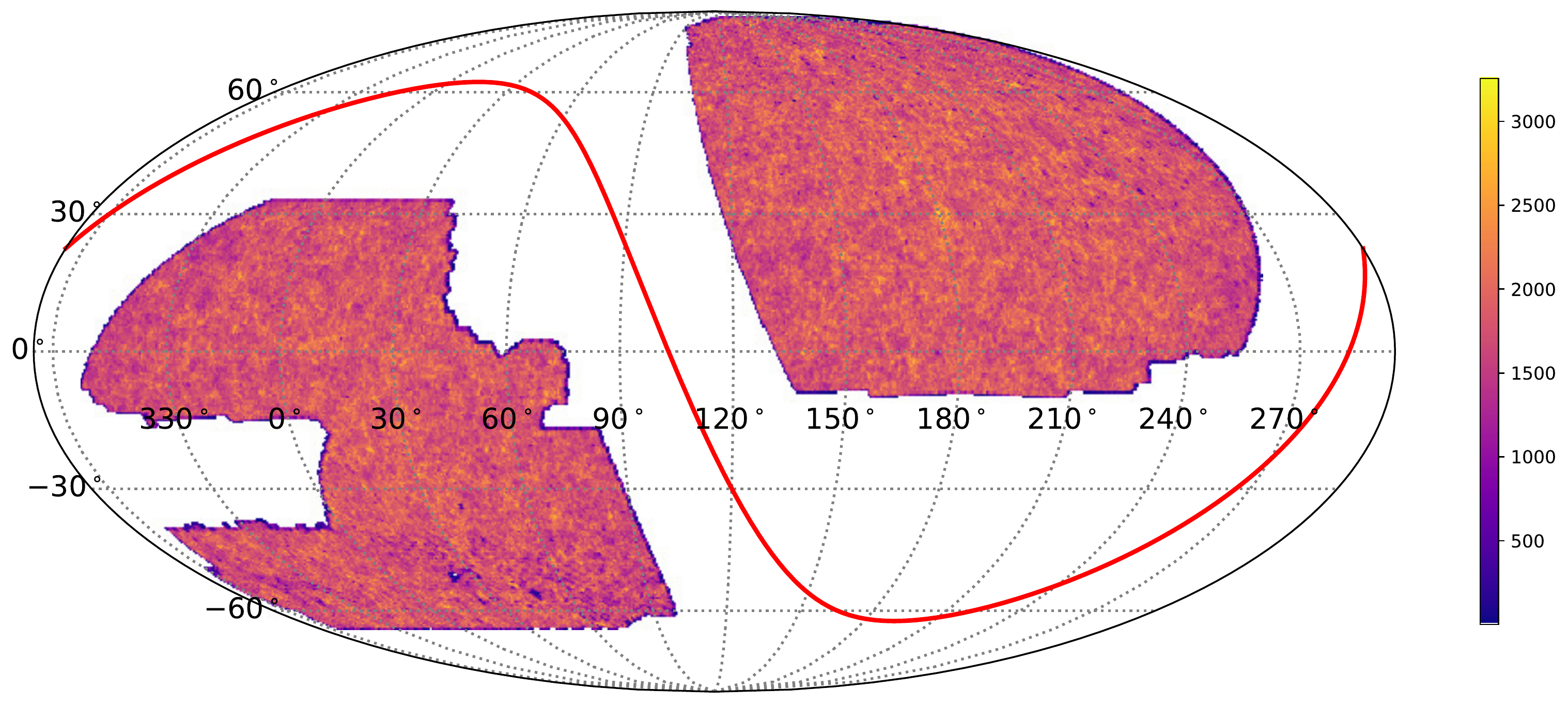}
\caption{The sky coverage of galaxies. These are drawn from two contiguous areas: the south galactic cap ($\sim$ 8580 square degrees with 59.6 million galaxies) and the north galactic cap ($\sim$ 9673 square degrees with 69.75 million galaxies). The color code is the galaxy number counts per 0.25 square degrees.}
\label{fig:sky_cov}
\end{figure*}

\subsection{DESI Legacy Imaging Surveys DR8}
\label{sec:DESIDR8}

The DESI Legacy Imaging Surveys provide the target catalogs for the DESI survey. The target selection requires a combination of three optical ({\it grz}) bands and two mid-infrared (W1 3.4$\mu$m, W2 4.6$\mu$m) bands. For the case of  galaxies, the {\it g/r/z} bands  have at least a 5$\sigma$ detection of 24/23.4/22.5 AB magnitude with an exponential light profile of half-light radius $0^{''}.45$.
The optical imaging data are provided by three public projects: the {\it Beijing-Arizona Sky Survey} (BASS), the {\it  Mayall z-band Legacy Survey} (MzLS), and the {\it  DECam Legacy Survey} (DECaLS), while the {\it Wide-field Infrared Survey Explorer} (WISE) satellite provides the infrared data. See an overview of the surveys in \citet{Dey2019}.

We select our sample from the latest public data release of the Legacy Surveys, the Data Release 8 (DR8). DR8 also includes data from some non-DECaLS surveys, with the majority originating from the {\it Dark Energy Survey} (DES). The catalogs are processed by the software package THE TRACTOR \citep{Lang2016} for source detection and optical photometry. 

Our galaxy samples first select those object with morphological classification as type {\it REX}, {\it EXP}, {\it DEV} and {\it COMP} from Tractor fitting results\footnote{In the Tractor fitting procedure, {\it REX} stands for round exponential galaxies with a variable radius,  {\it DEV} for deVaucouleurs profiles (elliptical galaxies), {\it EXP} for exponential profiles (spiral galaxies), and {\it COMP} for composite profiles that are deVaucouleurs plus exponential (with the same source center).}.

Following suggestions from the DESI Target selection (also see \citealt{Zou2019, Zhou2020, Ruiz-Macias2020,Zhou2020b,Raichoor2020,Yeche2020}), we require objects to have at least one exposure in each optical band and remove the area within ${\rm |b| < 25.0^{\circ}}$ (where ${\rm b}$ is the Galactic latitude) to avoid high stellar density regions. In addition to the Tractor morphology selection, a minimum data
quality flags are employed to remove the flux contaminations
from nearby sources (FRACFLUX) or masked pixels (FRACMASKED):
\begin{equation} \label{eqn:quality}\nonumber
\begin{aligned}
    \text{FRACMASKED}\_i < 0.4 \\
    \text{FRACIN}\_i > 0.3\\
    \text{FRACFLUX}\_i <0.5
\end{aligned}
\end{equation}
where $i \equiv g,r,z$. FRACIN is to select the objects
for which a large fraction of the model flux is in the
contiguous pixels where the model was fitted.
We also remove any objects close to bright stars by masking out objects with the following bits number in DR8 {\it MASKBITS} column: 1 (close to Tycho-2 and GAIA bright stars), 8 (close to WISE W1 bright stars), 9 (close to WISE W2 bright stars), 
11 (close to fainter GAIA stars), 12, and 13 (close to an LSIGA large galaxy and globular cluster, respectively). 
Note that all the magnitudes used in this paper are in the AB system and have been corrected for Galactic extinction by using the Galactic transmission values provided in DR8.

We adopt the random forest algorithm based photometric redshift estimation for the galaxies in our sample from the {\it Photometric Redshifts for the Legacy Surveys} ({\it PRLS} \footnote{https://www.legacysurvey.org/dr8/files/\#photometric-redshift-files-8-0-photo-z-sweep-brickmin-brickmax-pz-fits}, \citealt{Zhou2020}). To ensure the photo-z quality in the sample, we further limit our analysis to galaxies with ${\rm z\le 21}$ 
(above which the photometric redshifts are not reliable) and $ 0 < $ z\_phot\_median $\le 1$, where the z\_phot\_median is the median value of photo-z probability distribution function (PDF). See the details regarding the photometric redshift training and performance in appendix B of \citet{Zhou2020}. To make the sample's redshift information as accurate as possible, we replace the photometric redshift with the spectroscopic redshift, if available.  In total, there are 2.1 million galaxies with spectroscopic redshifts in our sample.  The complete 
spectroscopic redshifts sources in the original catalog include: BOSS, SDSS, WiggleZ, GAMA, COSMOS2015, VIPERS, eBOSS, DEEP2, AGES, 2dFLenS, VVDS, and OzDES\footnote{Check the selection conditions in these surveys
in section 3.2, \citealt{Zhou2020}.}.  Additional spectroscopic redshifts were obtained by matching galaxies in our sample
with those in 2MASS Redshift Survey \citep[2MRS;][]{Huchra2012}, 6dF Galaxy Survey Data Release 3 
\citep[6dFGRS;][]{Jones2009}, and 2dF Galaxy Redshift Survey \citep[2dFGRS;][]{Colless2001}.

After applying all the above selection criteria, we have a total number of 129.35 million galaxies remaining in our sample, within which 69.75 million are located in the north galactic cap (NGC) and 59.6 million in the south galactic cap (SGC). The total sky coverage of the galaxy sample is 18,253 square degrees (9673 square degrees in NGC, 8580 square degrees in SGC). As an illustration, we show the final footprint of galaxies on the sky in Figure~\ref{fig:sky_cov}.

For each galaxy in our NGC and SGC samples,  according to the typical photoz errors in the Legacy Surveys DR8 photoz galaxy catalog \citep[e.g.][]{Zhou2020,Wang2020}, we assign  it with a photometric redshift error,
\begin{equation} \label{eq:sigma_photo}
\sigma_{\rm photo} = 0.01+0.015z\,.
\end{equation}
If a galaxy has a spectroscopic redshift, we set $\sigma_{\rm photo}=0.0001$. This value will be
used in our modified group finder. 

\begin{figure}[!]
\center
\includegraphics[height=6.0cm,width=8cm,angle=0]{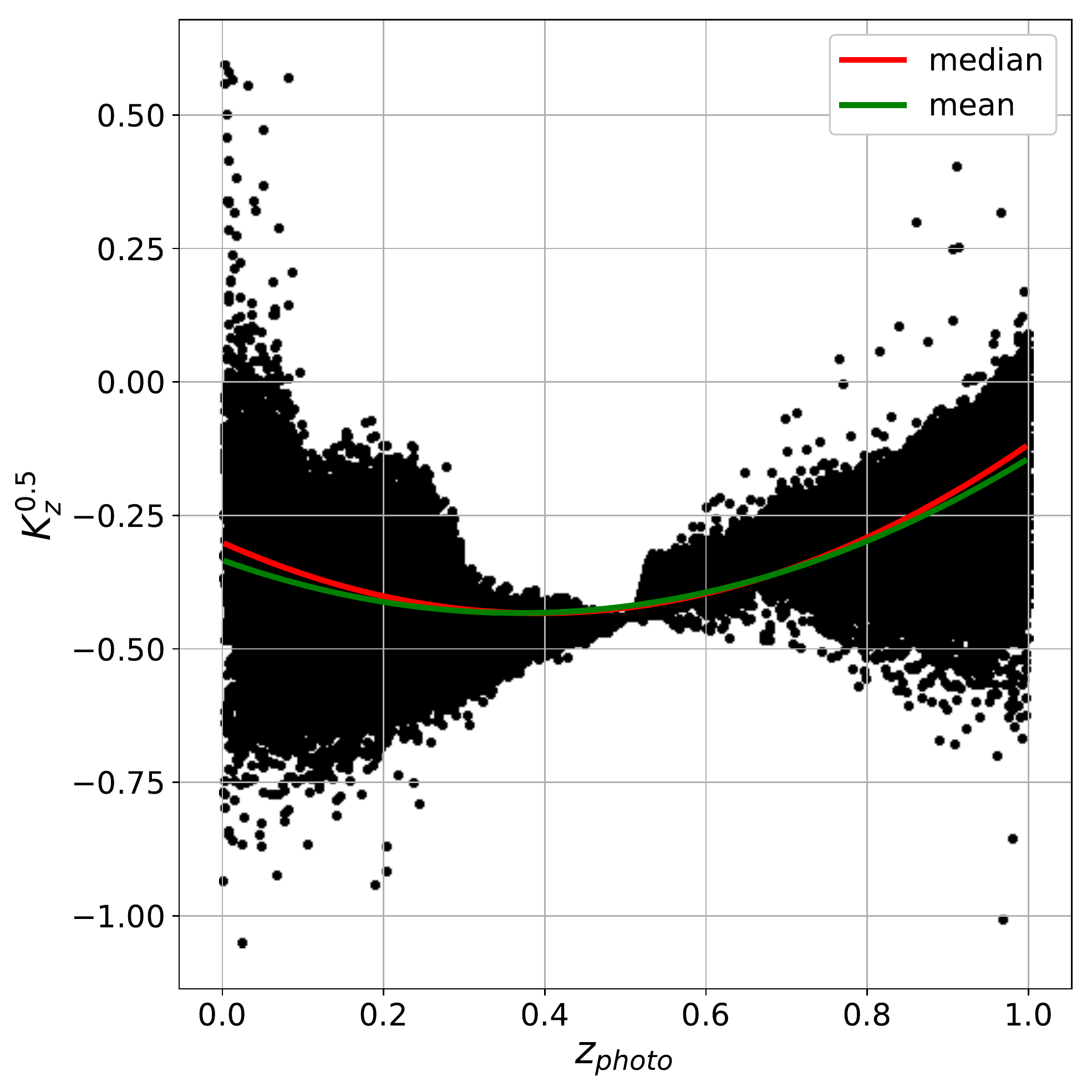}
\caption{ ${\rm K}$-correction to sample median redshift 0.5.
We show a random subset (50,000) of the sample galaxies, with the red (green) line fitting to the median (mean)
${\rm K}$-correction in redshift bins of 0.01. 
}
\label{fig:kcorrection}
\end{figure}

Next, for each galaxy, we use the following function to convert 
apparent magnitude to absolute magnitude according to its redshift,
\begin{equation} \label{eq:magnI}\nonumber
{\rm M}_{z} - 5\log h = 
{\rm m}_{z}  - {\rm DM}(z) - {\rm K^{0.5}_z}\,
\end{equation}
where DM($z$) is the distance module corresponding to redshift $z$, 
\begin{equation} \label{eq:distmeas}\nonumber
{\rm DM}(z) = 5 \log D_L(z) + 25
\end{equation}
with ${\rm D_L}$($z$) being the luminosity distance in ${\rm \mpch}$.
${\rm K^{0.5}_z}$ represents the ${\rm K}$-correction in z-band to sample median redshift $z\sim0.5$, obtained from the `Kcorrect' model (eg. v4\_3) described in \citet{Blanton2007}. For illustration, Fig. \ref{fig:kcorrection} shows  
${\rm K^{0.5}_z}$ values for a subset of our sample galaxies. 

In general, the average ${\rm K}$-correction can be described by the following function, 
\begin{equation} \label{eq:Kcorr}
{\rm K^{0.5}_z}(z) = az^2+bz+c\,.
\end{equation}
The fits to the median ${\rm K^{0.5}_z}$ values in the redshift bin of 0.01 result is shown in Fig.\ref{fig:kcorrection} as the red line
with $\lbrace 0.86\pm{0.06}, -0.67\pm{0.06}, -0.3\pm{0.01}\rbrace$ for $\lbrace a,b,c \rbrace$, respectively. For the mean ${\rm K^{0.5}_z}$ values in the redshift bin of 0.01, the fitting result is $\lbrace 0.73\pm{0.05}, -0.54\pm{0.05}, -0.33\pm{0.01}\rbrace$ for $\lbrace a,b,c \rbrace$. The fits agrees quite well with the overall ${\rm K}$-correction trend.  We will use the best fitting results to the 
median ${\rm K^{0.5}_z}$ values to model the K-correction of galaxies in our mock redshift surveys, as well as in calculating the maximum redshift a galaxy can be observed in the next subsection. 

The luminosity of each galaxy is then calculated using the following formula: 
\begin{equation} \nonumber
\log_{10} L = 0.4*(4.80-{\rm M}_z)\,
\end{equation}
with taking K-correction of the sun, ${\rm K^{0.5}_z}(0.0)=-0.3$,  into account,
to be consistent with ${\rm M_\odot}$ being 4.5 in z-band \citep{Willmer2018}.

\begin{figure*}
\center
\includegraphics[height=12.0cm,width=12cm,angle=0]{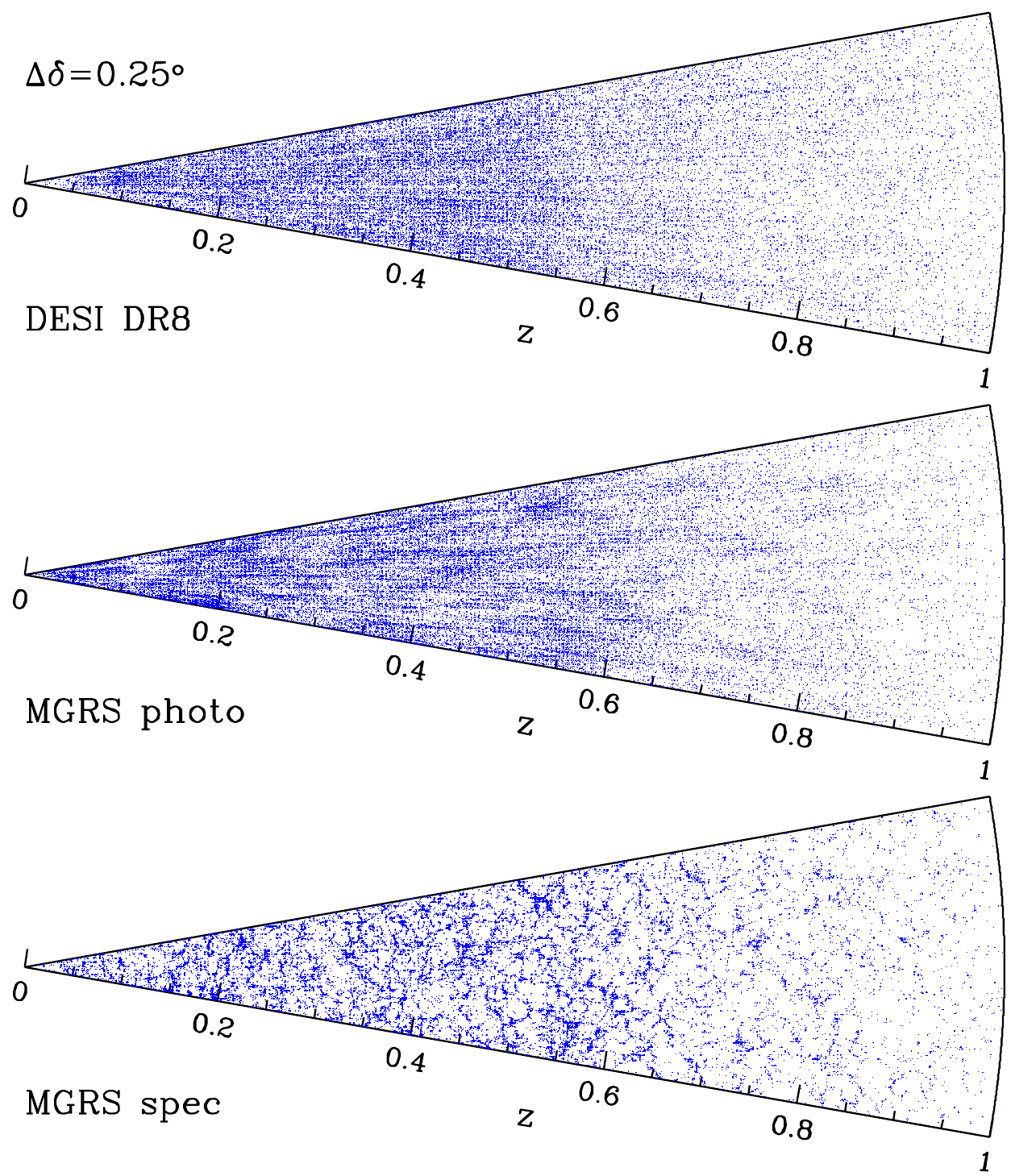}
\caption{Lower panel:  the projected distribution of a small subset of galaxies  along the declination direction (with $\Delta\delta=0.25^o$) in the mock spectroscopic redshift survey. Middle panel: same as the lower panel, but for the mock photometric redshift survey, where galaxies are added with additional photometric redshift errors specified by Eq. \ref{eq:sigma_photo}. Upper panel: the projected distribution of a subset of galaxies in the Legacy Surveys DR8. }
\label{fig:mgrs}
\end{figure*}

\begin{figure}
\center
\includegraphics[height=6.0cm,width=8cm,angle=0]{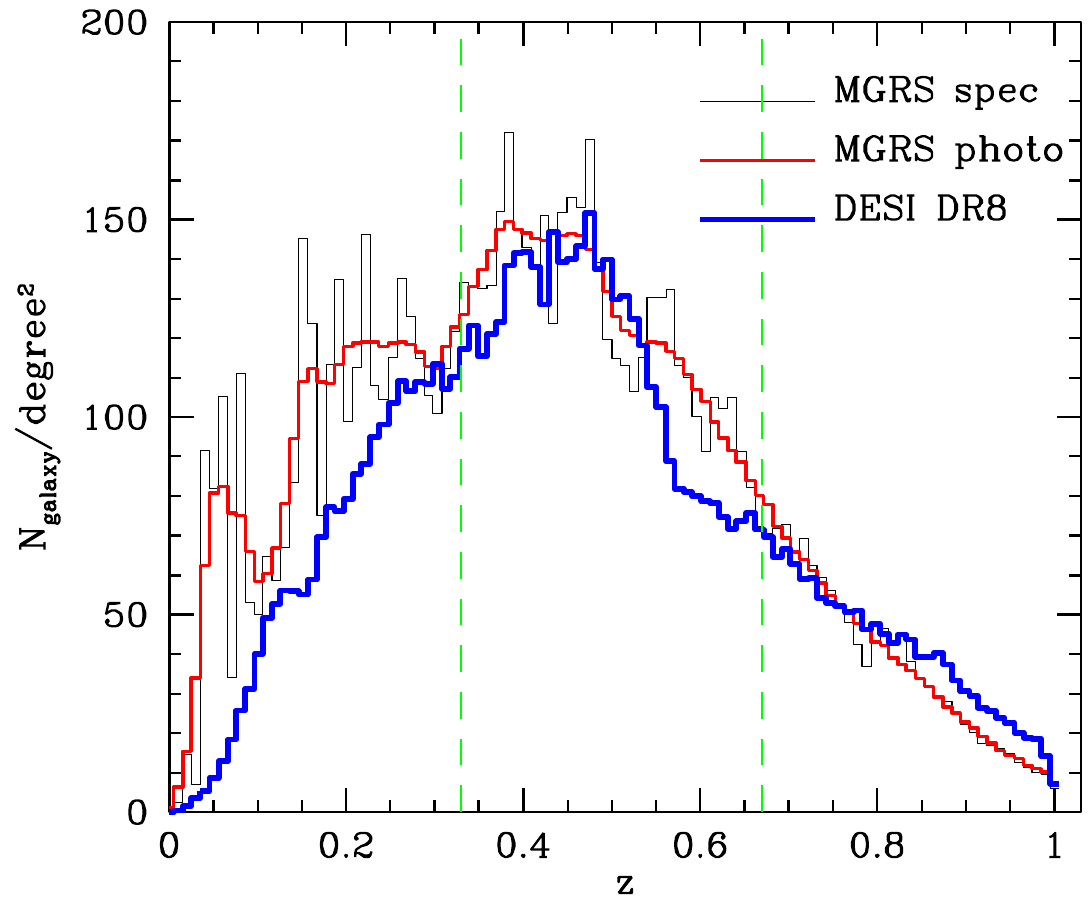}
\caption{The redshift distribution of galaxies in our MGRS and DESI DR8 samples, where the number of galaxies is normalized with respect to the sky coverage of the sample. The galaxy samples are separated into three redshift bins as indicated by the vertical dashed lines. }
\label{fig:zdist}
\end{figure}

\begin{figure*}
\center
\includegraphics[height=13.0cm,width=13cm,angle=0]{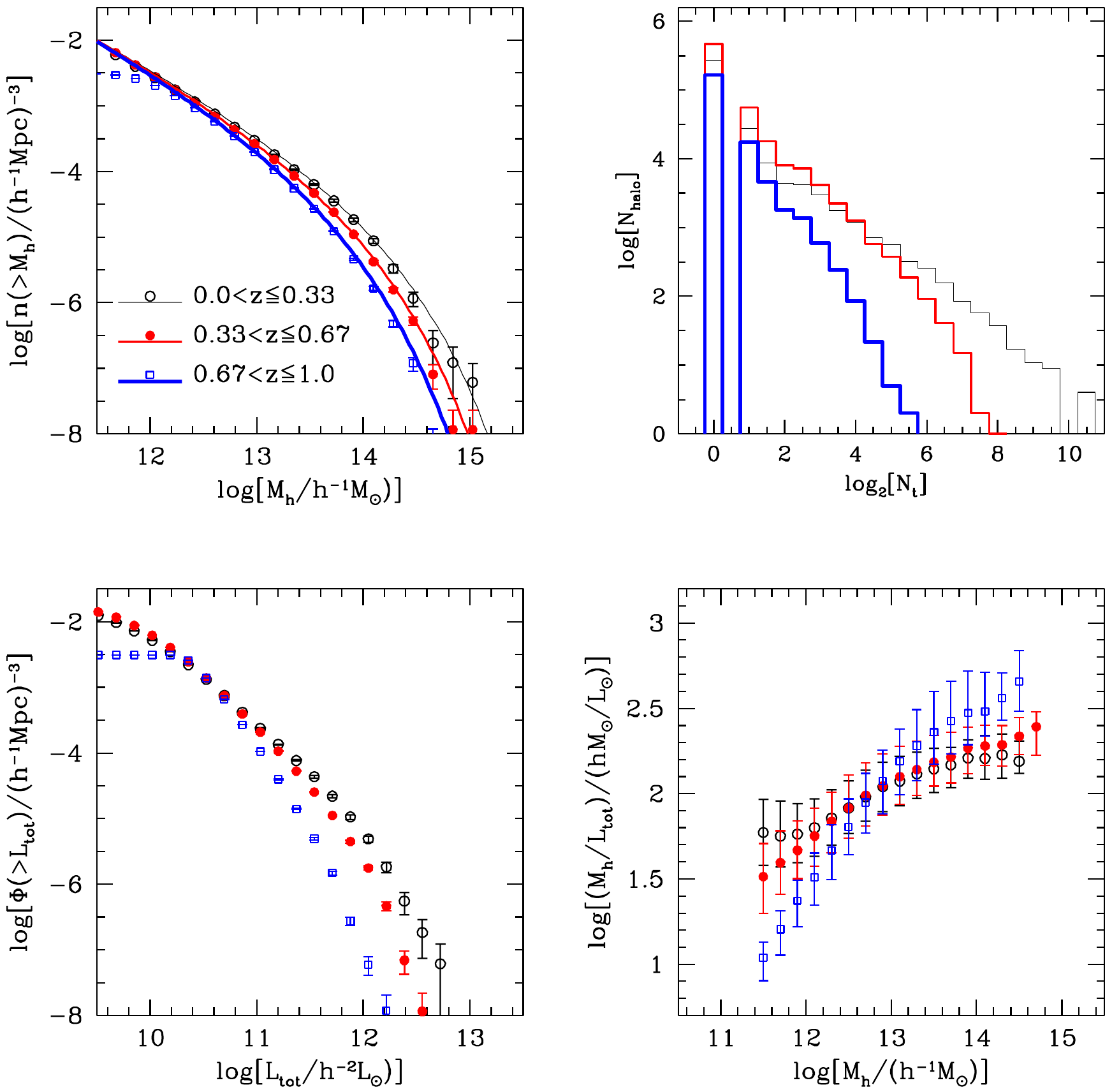}
\caption{Upper left panel: the {\it accumulative} halo mass functions of groups in different redshift bins as indicated  (same color coded for all panels in this figure). The symbols with errorbars are the results measured from the MGRS. The solid lines represent the corresponding theoretical model predictions given by SMT2001 for a WMAP5 cosmology, which match by construction. Upper right panel: the richness distribution of groups in the MGRS. Lower left panel: the {\it accumulative} group luminosity function, where group luminosity is the sum of luminosity of all member galaxies with apparent magnitude ${\rm z\le 21}$ in the same halo. Lower right panel:  the median mass-to-light ratio of groups as a function of halo mass in the MGRS. All errors obtained in this plot are estimated using 200 bootstrap resamples.}
\label{fig:mgLF}
\end{figure*}

\subsection{Simulation and mock galaxy redshift survey}
\label{sec:MGRS}

The N-body simulation used in our study is the ELUCID simulation carried out at the High Performance Center (HPC) at the Shanghai Jiao Tong University \citep{Wang2016}.  ELUCID evolves the distribution of $3072^{3}$ dark matter particles in a periodic box of $500 \mpch$ (${\rm L_{500}}$) from redshift $z=100$ to $z=0$. The simulation was run with the code {\tt L-GADGET}, a memory-optimized version of {\tt GADGET2} \citep{Springel2005}. The cosmological parameters adopted by ELUCID are consistent with the WMAP5 results and are the following:
$\Omega_{\rm m} = 0.258$, $\Omega_{\rm b} = 0.044$, $\Omega_\Lambda = 0.742$, $h =0.72$, $n_s = 0.963$ and $\sigma_8 = 0.796$ while each particle has a mass of $3.0875\times10^{8}\msunh$. The simulation is constrained (in terms of initial conditions) by the re-constructed initial density field of SDSS DR 7 \citep{Elucid2014}. ELUCID has been already used to study galaxy quenching \citep{Wang2018Elu}, galaxy intrinsic alignment \citep{Wei2018elu}, and cosmic variance \citep{Chen2019elu} while it was also combined with the abundance matching method to evaluate galaxy formation models \citep{Yang2018} and the L-Galaxies semi-analytic model \citep{Katsianis2020b}. 

Dark matter halos were identified with the standard Friend-of-Friends (FoF) algorithm \citep{Davis1985} with a link length equal to 0.2 times the average particle separation and with at least $20$ particles. The halo mass is calculated by summing up the mass of all the particles linked to the dark matter halo. As shown in \citet{Wang2016}, the halo mass function obtained from this simulation is in good agreement with the analytic model prediction by \citet{Sheth2001} (SMT2001) at the mass ranges of interest.  Based on halos at different outputs, halo merger trees were constructed \citep[see][]{Lacey1993}.  A halo in an earlier output is considered to be a progenitor of a present halo if more than half of its particles are found in the present halo.  The main branch of a merger tree is defined as all the progenitors going through from the bottom to the top, choosing always the most massive branch at every branching point. These progenitors are labeled as the main-branch progenitors.  Based on the above merger trees, galaxy catalogs were  generated using the empirical galaxy formation model described in  \citet{Chen2019elu}. Note that by combing the true and Monte Carlo merger trees, we have generated galaxies in all the halos in our ELUCID simulations with mass $\ge 10^9\msunh$, with a minimum galaxy luminosity  below $10^8\Lsunhh$ in all the simulation snapshots that we are going to use.

In this study, we create a mock galaxy redshift survey (MGRS) from the 
above-constructed galaxy catalogs. 
The MGRS,  with z-band galaxy magnitude ${\rm z\le 21}$, spans a redshift range of $0.0<z\le1.1$,  maximum angles of $\Delta \delta=10.0^o$, $\Delta \alpha=20.0^o$ and thus a total sky coverage at $199.74 \deg^2$. Here
we have taken into a negative K-correction to obtain the apparent magnitude for each galaxy.  Since the redshift range we are
interested in, $0.0<z<1.1$, is quite large, we have taken into
account the light cone effect by stacking simulation boxes at different redshifts
that are most consistent with the redshifts where they locate. 
The details of making the MGRS is described in \citet{Meng2020}. 

On top of the redshift information from the mock galaxy catalog,  we assign each galaxy a random shift drawn from a Gaussian distribution with dispersion $\sigma_{\rm photo}*(1+z)$, where $\sigma_{\rm photo}$
is the typical photoz errors in the Legacy Surveys DR8 photoz galaxy catalog described by Eq.\ref{eq:sigma_photo}. This value is also kept in the MGRS 
for our modified group finder to use.   We select galaxies with redshifts $0.0<z\le1.0$ from the mock galaxy catalog. There are a total of $1681566$ and $1679924$ galaxies remaining in our mock spectroscopic and photometric redshift surveys, respectively.  As an illustration, we show the projected distribution of a small subset of galaxies  along the declination direction (with $\Delta\delta=0.25^o$) within redshift $0.0<z\leq1.0$  in Fig. \ref{fig:mgrs},  where the lower and middle panels show our results for the mock spectroscopic and photometric redshift surveys, respectively. Compared to the clear cosmic web structures shown for the spectroscopic MGRS, galaxies in the photometric MGRS have more stretched structures along the line-of-sight. For comparison, we present the distribution of a small subset of galaxies in the Legacy Surveys DR8 in the upper panel of Fig. \ref{fig:mgrs}. The overall features in the photometric MGRS and the Legacy Surveys DR8 are quite similar. 

Once the MGRS is generated, we provide some intrinsic properties of the groups/clusters. 
Taking into account the fact that the halo mass functions and group/cluster luminosity functions may change significantly in the redshift range we are probing, we separate galaxies/groups/clusters in our MGRS (as well as our observed galaxy samples) into three redshift bins: $0.0<z\le0.33$, $0.33<z\le0.67$, $0.67<z\le1.0$. As an illustration, we show in Fig. \ref{fig:zdist} the redshift distribution of galaxies in our MGRS and DESI DR8 samples, where the number of galaxies is normalized with respect to the sky coverage of the sample. The vertical dashed lines separate galaxies in our samples into the three redshift bins. 

Shown in the upper left panel of Fig.\ref{fig:mgLF} are the {\it accumulative} halo mass functions of groups/clusters in different redshift bins as indicated. Symbols with errorbars are derived from the MGRS. 
Note that in order to partially reduce the impact of magnitude cut $z\le 21$, we separate the galaxy samples into $N_{\rm zbin}=3$ redshift bins and 
use a $V_{\rm max}$ method to calculate the halo mass functions,
\begin{equation} \label{eq:haloMF}
n(>M_h) = \sum_i 1/V_{\rm max} \,,
\end{equation}
where, $i$ is the ID of the halo/group, $V_{\rm max}$ is calculated according
to the apparent magnitude of the brightest group galaxy (BGG) at which redshift, $z_{\rm max}$, it can be
observed,
\begin{equation} \label{eq:V_max}
V_{\rm max}=V({\rm min}[z_{\rm max},z_{\rm bin, up}]) - V(z_{\rm bin, low})\,.
\end{equation}
Here $z_{\rm bin, low}$ and $z_{\rm bin, up}$ are the lower and upper limits of the corresponding redshift bin, respectively. In calculating the $z_{\rm max}$ of a galaxy, we have properly taken into account the difference in the K-correction between the actual and maximum redshifts, $z_{\rm max}$, using Eq. \ref{eq:Kcorr}.
As a comparison, we also show the corresponding theoretical model predictions given by SMT2001 represented by the solid lines.
We demonstrate that overall the model and data agree quite well, though some small deviation shown at the high mass end for the low redshift bin, which is mostly caused by the small volume of our sample. 
On the other hand, the deviation at the low mass end for the high redshift bin, although we have partly corrected using the $V_{\rm max}$ method, is due to the survey magnitude limit cut, $z\le 21$, where faint galaxies residing in low mass halos can not be observed. In the upper right panel, we present the richness distribution of groups/clusters in the MGRS. In the lowest redshift bin, there are only a few groups with more than 1000 members, while in the high redshift bin, the richest groups host around $30$ members. At the lower left panel, we show the {\it accumulative} total luminosity function of groups/clusters, with the total luminosities $L_{\rm tot}$ calculated by summing up all the member galaxy luminosities with apparent magnitude ${\rm z\le 21}$  in the MGRS.  Here again, a $V_{\rm max}$ method similar to Eq. \ref{eq:haloMF} is used to partly correct the group incompleteness. Obviously, at the high redshift bin, the groups/clusters with $L_{\rm tot}<10^{10.2}\Lsunhh$ are missing from our MGRS, which simply reflects the facts that galaxies in this redshift bin below this luminosity did not meet the magnitude limit cut. In the lower right panel, we present the median mass-to-light ratio of groups/clusters in the MGRS. Overall, the mass-to-light ratio in high mass  groups/clusters $\sim 10^{14.5}\msunh$ spans from about 200 at low redshift to about 500 at high redshift. In low mass groups with mass  $\sim 10^{11.5}\msunh$ the related numbers decrease to 50 to 10, respectively. Note that all the errorbars obtained in this figure are estimated using 200 bootstrap re-samplings.

\begin{figure*}
\center
\includegraphics[height=6.0cm,width=13cm,angle=0]{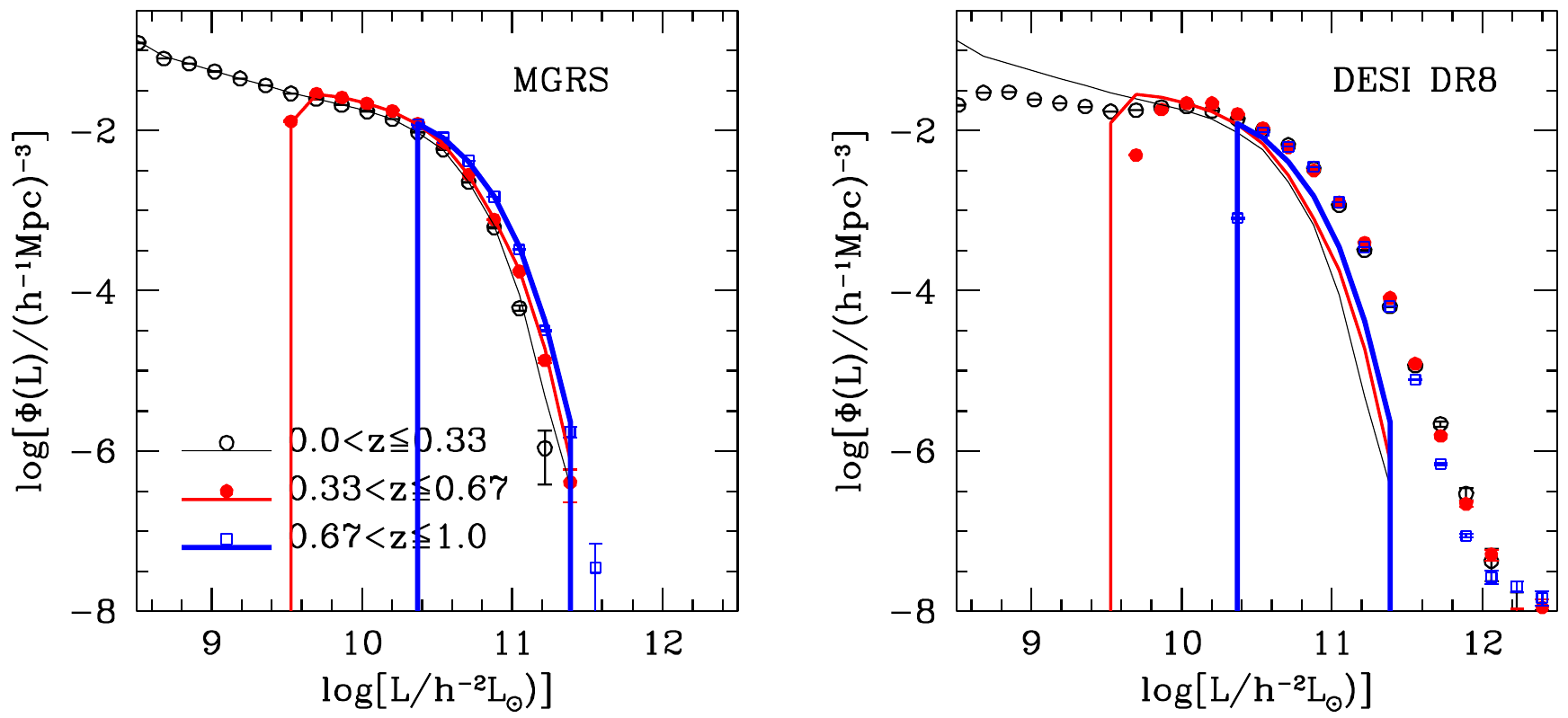}
\caption{Left panel: the galaxy luminosity functions
  obtained from galaxies in the MGRS. The data points with errorbars are the results obtained from galaxies with mock spectroscopic redshifts in different redshift bins as indicated. The solid lines represent results obtained from galaxies using the mock photometric redshifts correspondingly. Right panel: the galaxy luminosity functions obtained from galaxies in the Legacy Surveys DR8. The Symbols are for galaxies in different redshift bins as those in the left panel. The solid lines are the same as those shown in the left panel for comparison.}
\label{fig:zLF}
\end{figure*}

\subsection{The $z$-band galaxy luminosity functions}
\label{sec:zLF}

Before we evaluate the group/cluster finding algorithm with the mock catalogs and apply the method to the Legacy Surveys DR8 observational data, we measure the galaxy luminosity functions (LFs) from the mock and observational galaxy catalogs. The galaxy luminosity function, which describes the average light densities of the Universe, represents a fundamental tool in understanding galaxy formation and evolution. Since galaxies are formed within dark matter halos and the most massive (central) galaxies reside roughly in the most massive halos, there exists statistical relations between galaxy luminosities and halo mass \citep[e.g.][and references therein]{Yang2012}.  The accurate estimation of the galaxy luminosity functions plays a key role in establishing those kinds of relations.  We estimate the galaxy luminosity function by counting galaxies
in each luminosity bin and weighted by $1/V_{\rm max}$,
\begin{equation} \nonumber
\Phi(L) {\rm d} \log L= \sum_i 1/V_{\rm max} \,.
\end{equation}
Here $V_{\rm max}$ can be similarly calculated using Eq. \ref{eq:V_max}, but for the galaxy's own $z_{\rm max}$.

We first focus on galaxies with mock spectroscopic redshifts in the MGRS. Shown in the left panel of Fig. \ref{fig:zLF} are the $z$-band luminosity functions of galaxies in three redshift bins: $0.0<z\le0.33$, $0.33<z\le0.67$, $0.67<z\le1.0$, respectively. The galaxy luminosities are obtained from the $z$-band absolute magnitude $K$-corrected to redshift $z=0.5$. In general, we see that the LFs measured in different redshift bins are roughly consistent with each other at the bright end.  Since the MGRS is only complete up to ${\rm z\le 21}$, faint galaxies will not be observed, especially those at higher redshifts. Thus, we see that the derived LFs have suffered more from incompleteness problem in the higher redshift bins,  with brighter faint luminosity end cutoffs. Again, the errorbars are obtained from 200 bootstrap re-samplings of galaxies. 

Next, we focus on galaxies with photometric redshifts in the MGRS. The solid lines shown in the left panel of Fig. \ref{fig:zLF} are obtained using the mock photometric redshift galaxies. Overall, the results are in good agreement with those obtained from the mock spectroscopic redshift galaxy samples. Since in our mock photometric redshifts, we only add a small redshift error to each galaxy, without any catastrophic redshift assignment. That is the main reason that we do not see a significant Eddington bias which might be induced in most of the photometric redshift surveys. 

Finally, we focus on the Legacy Surveys DR8 data. The symbols shown in the right panel of Fig. \ref{fig:zLF} are the retrieved galaxy luminosity functions in three redshift bins.  Compared to the MGRS (shown as solid lines), galaxies in the Legacy Surveys DR8 show similar cutoffs at low luminosity ends, which is caused by the same $z\le 21$ magnitude cut. On the other hand, galaxies in the Legacy Surveys DR8  
   are somewhat brighter. This is mostly due to the galaxy luminosity calibration as well as the different K-correction modelings. Here we used the photometric data in COSMOS2015 catalog \citep{Laigle2016} to calibrate the luminosity of the galaxies in our catalog.  The galaxy luminosity is firstly assigned in $y$ band using a luminosity-stellar mass abundance matching method. Then a $z$ band galaxy luminosity is obtained according to the $z-y$ color of the galaxy, which is obtained using the age distribution matching method \citep{Hearin2013}. For each galaxy in our catalog, we choose a galaxy with the same luminosity and color in COSMOS2015 and use this galaxy's fitted spectrum to calculate the K-correction and hence the apparent magnitude of that galaxy. See \citet{Meng2020} for more details. Note that since the K-correction used in \citet{Meng2020} was corrected to redshift $z=0.0$, in order to be consistent with the one used in this paper, we use $z$-band apparent magnitude and Eq. \ref{eq:Kcorr} to calculate the absolute magnitude and luminosity of each galaxy in our MGRS. 
Nevertheless, since in our cluster finder, we are using the accumulate luminosity functions measured from the data in different redshift bins to estimate the halo mass, such systematic differences will not impact the group/cluster membership determination.

\section{The extended halo-based group finder}
\label{sec_steps}

The halo-based group finder developed 
in \citet{Yang2005a, Yang2007} has been tested and successfully applied to 
galaxy samples with spectroscopic redshifts (see 
references listed in the introduction).
The strength of the algorithm lies in its iteration nature and utilization of 
an adaptive filter while taking into account the general properties of dark matter halos. 
However, it can only be applied to galaxy samples with spectroscopic redshift.
In this work, 
we extend the algorithm so that it can deal with
galaxies with photometric and spectroscopic redshift simultaneously.

Below we list the main steps that are used in our extended group finder. Note that the basic framework of this extended halo-based group finder is quite similar to the one 
used in Y05 and Y07, except those minor changes, the  most important extension/upgrade in our extended group finder is related with Eq. \ref{eq:pz} in Step 4.

{\bf Step 0: Start.}
We start our search by assuming that each galaxy is a group candidate.

{\bf Step 1: Measure the accumulative group luminosity functions.}
 We first measure the total luminosity within each group,
\begin{equation}\label{eq:LG}
L_{G} = \sum_i L_i\,,
\end{equation}
where $i$ is the ID and $L_i$ is the luminosity of each member galaxy.
Then we calculate the accumulative group luminosity functions from the whole group sample similar to Eq. \ref{eq:haloMF}.
Note that since the galaxy/group samples from observations are usually flux
limited, measurements at different redshifts suffer from different absolute magnitude limits.  In order to partially reduce the impact of
this observational selection effect, here again we separate the galaxy samples into $N_{\rm zbin}=3$ redshift bins and 
use a $V_{\rm max}$ method to calculate the accumulative group luminosity functions (e.g. Eq. \ref{eq:V_max}).

{\bf Step 2: Determine the mass-to-light ratio of the groups.}  In each redshift bin, 
the accumulative halo mass function is calculated at the median redshift of the galaxies of interest. 
The mass-to-light ratios of the groups in each redshift bin are then computed using the abundance matching method according to the accumulative halo mass functions and group luminosity functions (see section 3.5 of Y07). {Here we have properly taken into account the incompleteness of the halos/groups as described in Section \ref{sec:comp2}. }

{\bf Step 3: Estimate the mass, size, and velocity dispersion of the dark matter halo associated with each tentative group.}  With the
mass-to-light ratios in different redshift bins determined in step 2, we assign each tentative group with a halo mass, $M_L$, by interpolations on both the redshift and total luminosity of the group.
  
In our study, we define dark matter halos as having an overdensity of $180$ times the background density of the universe. This corresponds to a comoving halo radius of
\begin{equation}\nonumber
r_{180} = 0.781 \mpch \left( \frac{M_L}{\Omega_m10^{14} \msunh}\right)^{1/3} \, 
\end{equation}
where $z_{\rm group}$ is the redshift of the group center. The line-of-sight velocity dispersion of dark matter particles is described by
\begin{equation}
\label{veldispfit}
\sigma_{180} = 632 \kms \left( \frac{M_L \Omega_m}{10^{14} 
\msunh}\right)^{0.3224}\,.
\end{equation}
This fitting function is obtained by \citet{vandenBosch2004} to describe the halo mass dependence of the one-dimensional velocity dispersion of dark matter particles using the halo concentrations of \citet{Maccio2007}.  Here slight modifications are made with respect to the ones using in Y07 so that they are 
applicable to $\Lambda$CDM cosmology with other $\Omega_m$ values. The best fitting values used in Eq. \ref{veldispfit} are obtained from a few sets of simulations with different $\Omega_m$ values (Frank C. van den Bosch, private communications).

{\bf Step 4: Determine the group memberships using halo information.}  We sort and start from the most massive group candidate to search its member galaxies using its halo properties. We assume that the distribution of galaxies in
phase-space follows that of the dark matter particles and the number density contrast of galaxies in the redshift space around the group
center (the luminosity weighted center) at redshift $z_{\rm group}$ can be written as
\begin{equation}\nonumber
P_M(R,\Delta z) = {\frac{H_0}{c}} {\frac{\Sigma(R)}{\bar \rho}} p(\Delta z) \,,
\end{equation}
where  $c$ is the  speed of  light, $\Delta  z =  z -  z_{\rm group}$, $\bar{\rho}$ is  the average density  of Universe, and  $\Sigma(R)$ is the projected surface density of  a (spherical) NFW \citep{Navarro1997} halo:
\begin{equation}\nonumber
\Sigma(R)= 2~r_s~\bar{\delta}~\bar{\rho}~{f(R/r_s)}\,.
\end{equation}
Readers who are interested in the detailed functional forms within this equation can go to  Y07 for more details.

%
%
%
%
The function $p(\Delta z){\rm d}\Delta z$ describes the redshift distribution of galaxies within the halo, and is assumed to have a Gaussian form,
\begin{equation}\label{eq:pz}
p(\Delta z)=  {1 \over  \sqrt{2\pi}} {c \over  \sigma (1+z_{\rm  group})} \exp
\left [ \frac {-(c\Delta z)^2} {2\sigma^2(1+z_{\rm group})^2}\right ] \,.
\end{equation}
Here we made our main modification by properly taking into account the photometric redshift error, as well as the galaxy velocity dispersion within dark matter halos, with $\sigma= \max (\sigma_{180}, c\sigma_{\rm photo})$ where $\sigma_{\rm photo}$ is the typical photometric redshift error listed in Eq. \ref{eq:sigma_photo}.
Note that in our galaxy samples, if a galaxy has a spectroscopic redshift, we assigned it with a $\sigma_{\rm photo}=0.0001$ value. For this galaxy, the $\sigma=\sigma_{180}$ value will automatically be used, while for the majority galaxies with only
photometric redshifts, $\sigma=c\sigma_{\rm photo}$.

With the ${\rm P_M(R,\Delta z)}$ defined above, we use the following procedure to decide whether a galaxy should be assigned to a
particular group. For each galaxy, we loop over all groups to compute the distance ${\rm (R,\Delta z)}$ to group center, where ${\rm R}$ is the projected distance at the redshift of the cluster.  If ${\rm P_M(R,\Delta z) \ge B{\sigma_{180} /\sigma}}$,  the galaxy will be assigned to the group.  As pointed out in Y05, the background value  ${\rm B}$ in perspective  is used as a
threshold for the redshift space  density contrast of groups. Ideally, for a spectroscopic redshift survey,
${\rm B}$ should correspond  roughly to the redshift space density
contrast at the edge of a halo, i.e.,
\begin{equation}\label{eq:B}
{\rm B} \approx {\rho_{\rm red}(r_{180}) \over \bar{\rho}} 
  \approx { \rho(r_{180}) \over \bar{\rho} }
  {(4\pi /3) r_{180}^3 \over \pi r_{180}^2 \sigma_{180} /H_0}\,. 
\end{equation}
Using that  ${\rho(r_{180}) / \bar{\rho}}\sim  30$ and ${\sigma_{180}  / H_0
  r_{180}} \sim 4$, we obtain ${\rm B} \sim 10$, which is quite independent of halo mass.
In this extended group finder, the related background value differs from the original ${\rm B}$ by a factor of 
${\rm \sigma_{180} / \sigma}$ to ${\rm B\sigma_{180} / \sigma}$, which accounts for the decrease of density contrast caused by the photometric redshift errors.

If a galaxy can be assigned to more than one group according to this criteria, it will be only assigned to the one for with the largest ${\rm P_M(R,\Delta z)}$ value. If all members in one group can be assigned to the other group, the two groups will be merged into a single group.

{\bf Step 5: Iterate.}  Using the new memberships obtained in Step 4, we re-compute the group centers and the group luminosity $L_G$ using Eq. \ref{eq:LG} and go back to Step 3. This
iterative process goes on until there is no further change in the group memberships. Once we finish all the group membership
assignment, we go back to Step 1 for another iteration. 
This iteration cycle stops as long as the mass-to-light ratios have converged,
which typically takes only 3 to 4 iterations.

In general, lowering the ${\rm B}$ value will increase both the completeness and contamination of group members. We note that in Y05 and Y07, the background value (in step 4) is set to ${\rm B=10}$
to balance the completeness and contamination of group members.
In this work since we aim to extend our algorithm to process photometric galaxy samples, 
where contamination is inevitable anyway, we focus more on improving the completeness of group/cluster members. {For this purpose, in addition to our fiducial ${\rm B=10}$ value, we will test the performance of 
our group finder by adopting a lower ${\rm B}$ value, e.g., ${\rm B=5}$.  According to Eq. \ref{eq:B}, this lower ${\rm B}$ value roughly corresponds to consider galaxies that are within 1.5-$\sigma$ photoz error (or velocity dispersion) regions along the line-of-sight. }

\begin{figure*}
\center
\includegraphics[height=13.0cm,width=13cm,angle=0]{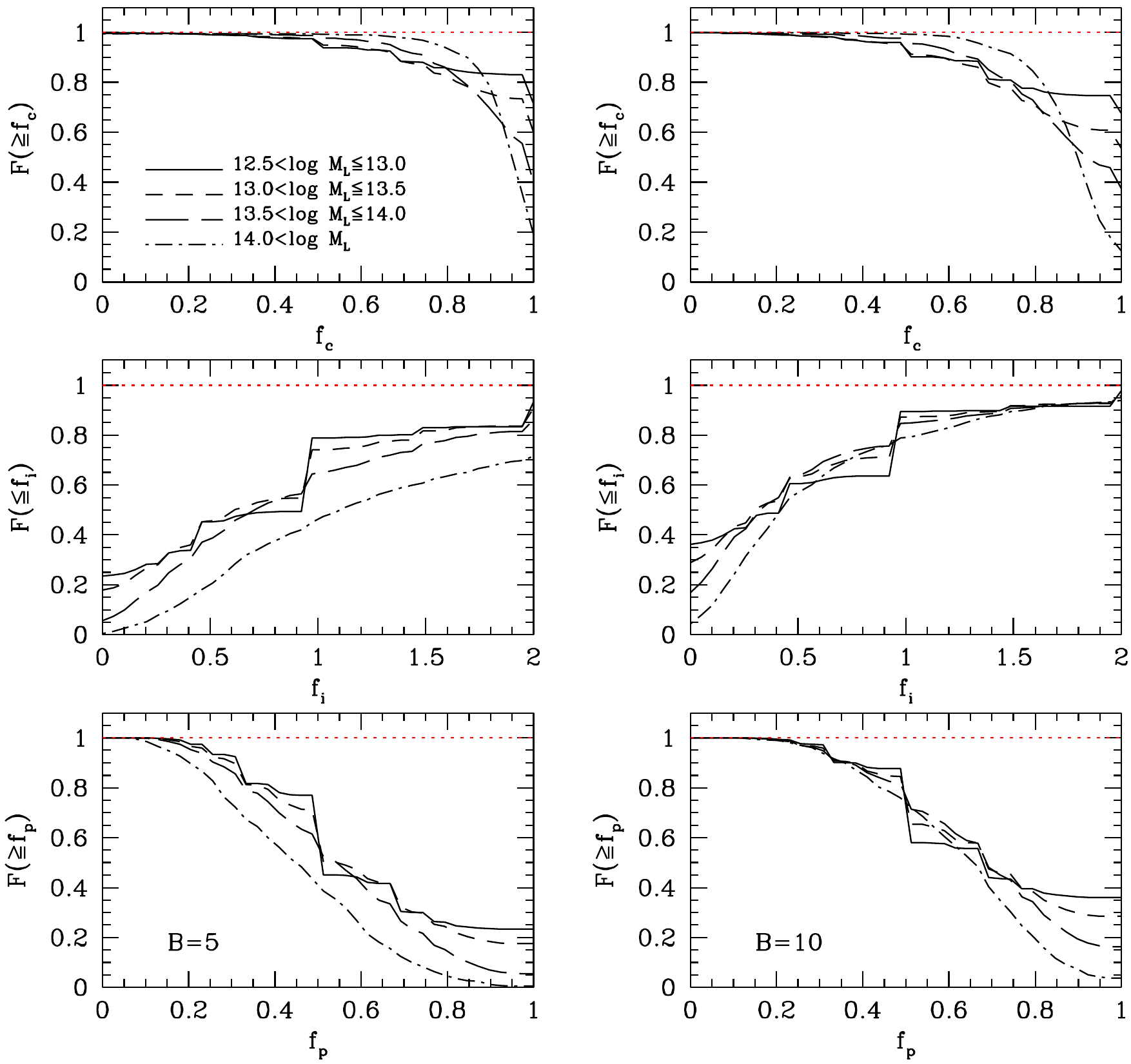}
\caption{The upper, middle and lower panels show the cumulative distributions of completeness,  $f_{\rm c}$  (the fraction  of selected true members), contamination, $f_{\rm i}$, (the fraction of interlopers), and purity,  $f_{\rm p}$,  (ratio between the selected true members  and the group members), respectively. See main text for the detailed definitions of all the above parameters.
The left and right panels show the parameters obtained for background values $B=5$ and $B=10$,  respectively.   Different lines show the result for groups with different assigned mass, $M_L$, as indicated.  Results are plotted for groups with at least 2 members, since groups with only 1 member have, by definition, $f_{\rm i}=0$. }
\label{fig:compl}
\end{figure*}

\section{Testing the performance of the extended group finder}
\label{sec_test}

To test the performance of our algorithm we run our extended halo-based group finder over the mock galaxy redshift survey (MGRS) as outlined in section \ref{sec:MGRS}. This MGRS mimics the depth and photometric redshift quality of the Legacy Surveys DR8.  

\subsection{Completeness, Contamination, and Purity of the group memberships}
\label{sec:performance}

\begin{figure*}
\center
\includegraphics[height=12.0cm,width=13cm,angle=0]{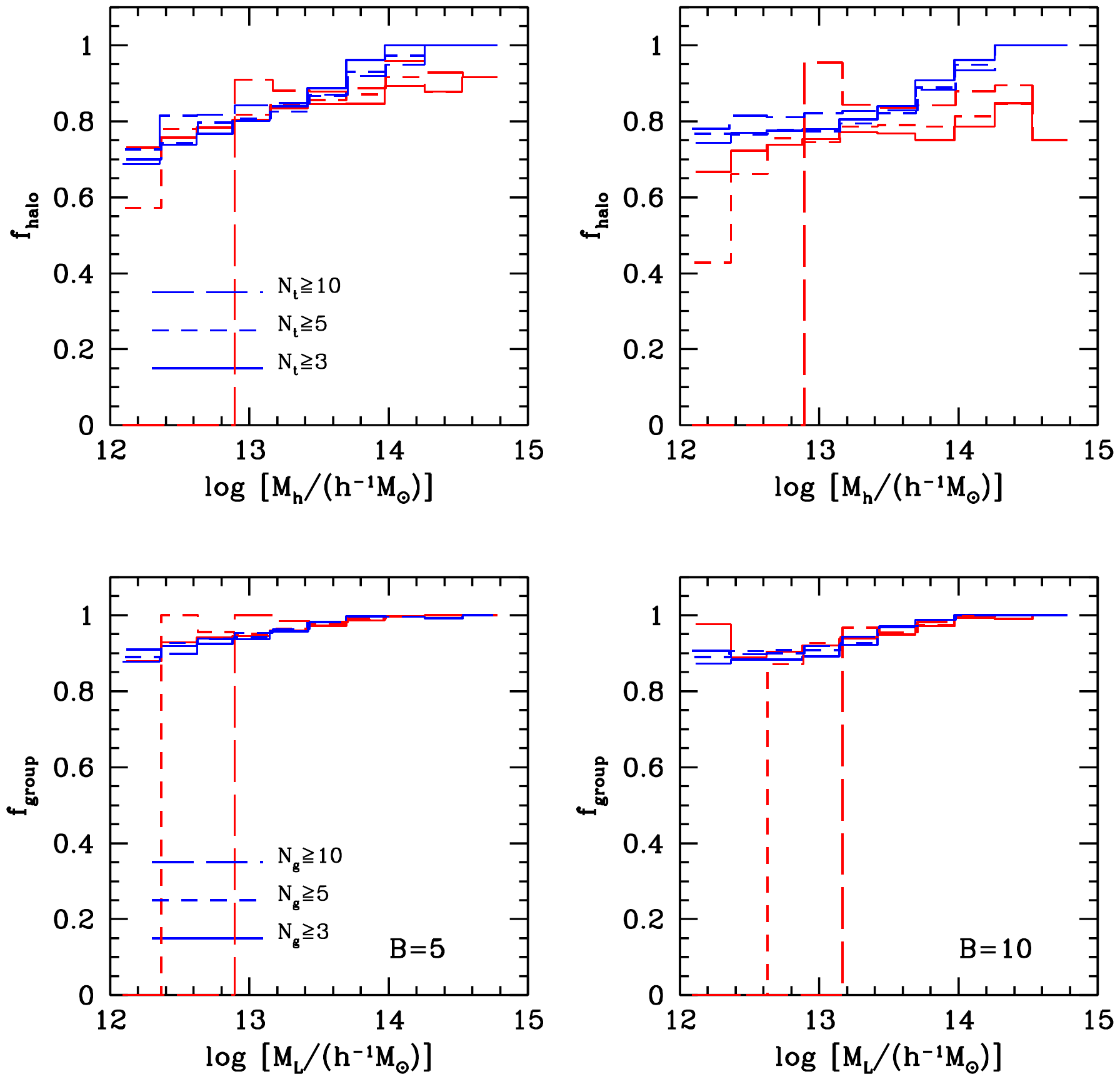}
\caption{Upper panels: The global halo completeness, $f_{\rm halo}$, defined as the fraction of halos  in the MGRS with more than 50\% members being identified as the main part of the group, as a function  of the true  halo mass  ${\rm M_h}$. Results are shown for 
  halos with  at least 3, 5 and 10 true members in the MGRS as indicated, respectively. And we use red and blue histograms to represents results for groups within two redshift bins, $z>0.5$ and $z\le 0.5$, respectively. Lower panels: the group purity, $f_{\rm group}$, defined as the fraction of identified groups which contain more than 50\% members of a true halo as a function  of the assigned halo mass ${\rm M_L}$. Results are shown for groups with  at least 3, 5 and 10 identified members, respectively. Here again, we use red and blue histograms to represents results for groups within two redshift bins, $z>0.5$ and $z\le 0.5$, respectively. The left and right  panels highlight the results obtained for background values of ${\rm B=5}$ and ${\rm B=10}$,  respectively. }
\label{fig:comp_halo}
\end{figure*}

To assess the performance of the group finder we proceed as follows: For a group ${\rm k}$, we investigate the halo ID for each member galaxy, and sum up the number of members in each halo. We choose the one with the most members as the candidate halo ${\rm h_k}$. Since in the photometric data, the brightest central galaxy (BCG) in a halo may move far away from the other members, we do not use the BCG to identify the candidate halo of the group, which is used in Y05. We define $N_{\rm t}$ as the total number of true members in the MGRS that belong to a halo ${\rm h_k}$, ${\rm N_s}$ is defined as the number of these true members that are selected as members of group  ${\rm k}$, ${\rm N_i}$ is defined as the number of interlopers (group members that belong to a different halo) and $N_{\rm g}$ is defined as the total number of selected group members. 
These values allow us to define, for each group, the following three quantities:
\begin{itemize}
\item The completeness $f_{\rm c} \equiv N_{\rm s}/N_{\rm t}$, 
\item The contamination $f_{\rm i} \equiv N_{\rm i}/N_{\rm t}$,
\item The purity $f_{\rm p} \equiv N_{\rm s}/N_{\rm g}$.
\end{itemize}
Note that since the contamination is defined with respect to the true number of group members, the  contamination $f_{\rm i}$ can be larger than unity. A contamination $f_{\rm i} > 1$ implies that the number of interlopers is larger than the number of true members being detected. An ideal group finder yields groups with $f_{\rm c} = f_{\rm p} = 1$ and $f_{\rm i}=0$. 

The results obtained from the MGRS are shown in Fig.~\ref{fig:compl} with the left panels being for ${\rm B=5}$ and the right panels for ${\rm B=10}$. Since groups with  a single  member  have  zero  contamination ($f_{\rm  i}=0$)  by definition, results are only shown for groups with a  richness of $N_{\rm g}\ge 2$.  The upper left-hand panel of Fig.~\ref{fig:compl}  shows the cumulative distributions of the  completeness  $f_{\rm   c}$ for the case of ${\rm B=5}$.  Different line-styles correspond  to  groups of  different  assigned halo  mass, $M_L$, as  indicated. There are about 83\% groups with group member completeness larger than 80 percent while  $\sim 93\%$ of groups have member completeness larger than 60 percent for groups with mass $\ga 10^{12.5}\msunh$. Among them, groups in the most massive bin shows the highest completeness. 
The middle left-hand panel of Fig.~\ref{fig:compl} shows the cumulative distributions of the contamination $f_{\rm i}$. 
{On average, around $70\%$ of the groups in mass range $10^{12.5}< \log M_L \le 10^{14.0}\msunh$ have $f_{\rm  i}\le 1$. While in the most massive bin, $10^{14.0}\msunh< \log M_L$, this number decreases to $43\%$. }
{Since there are $\sim 30\%$ of the groups with more interlopers than true numbers ($f_{\rm  i}>1$),}
special care needs to be taken
if one wants to make use of the member galaxies, which we will look into in a subsequent paper. Finally, the lower left-hand panel shows the cumulative  distributions of the purity $f_{\rm p}$. 
{On average there are about 70\% groups with mass $\ga 10^{12.5}\msunh$ (except those in the most massive bin) having purity $f_{\rm p}>0.48$. In the most massive bin, $10^{14.0}\msunh< \log M_L$, this number is about $41\%$. }

Next, we compute the completeness, contamination, and purity in terms of group membership for our fiducial case of ${\rm B=10}$. The corresponding results are  shown  in  the  right-hand panels  of Fig.~\ref{fig:compl}. Compared to the results shown in the left-hand panels for the case of ${\rm B=5}$, for groups  with mass {$10^{12.5}< \log M_L $} the performances differ in the following ways:

\begin{itemize}
\item (1) The member completeness is slightly lower. About 72\% groups are exhibiting more than 80 percent group member completeness,  and $\sim 90\%$ ($\sim 95\%$) groups have member completeness larger than 60 (50) percent.
\item (2) The member contamination in each group is smaller than that in the ${\rm B=5}$ scenario. Here about $85\%$ of groups have an interloper fraction  $\le 1$.
\item (3) The membership purity in each group is also better. About 83\% groups  having purity $f_{\rm p}>0.48$.
\end{itemize}

{Comparing groups in different mass bins, we notice that groups in the highest and lowest mass bins have better completeness and worse purity than groups in the intermediate mass bins.  For the case B=10, the differences between mass bins are slightly less pronounced. }

Given the differences between the ${\rm B=10}$ and ${\rm B=5}$ cases, if we care more about the purity of the true group members, the case ${\rm B=10}$ is preferred.

\subsection{Global completeness and purity of the groups}
\label{sec:comp2}

In this subsection, we focus on the global properties of the groups we find from the MGRS. 
To further evaluate the performance of our extended halo-based group finder, 
one useful quantity we want to assess is the {\it  global completeness} of groups, $f_{\rm halo}$, defined as the fraction of halos in the MGRS with more than half of the true members being identified by the group finder as their richest components. 
The upper left panel of Fig.~\ref{fig:comp_halo} shows $f_{\rm halo}$ obtained from our MGRS for halos with $N_{\rm t} \geq 3, 5, 10$ as functions of the true halo mass for the ${\rm B=5}$ case. Note that since in our MGRS, the photoz error depends quite strongly on redshift (see Eq. \ref{eq:sigma_photo}), and because of the $z\le 21$ magnitude cut, groups at higher redshift should have less members and suffer more from the photoz errors. To check the related impact, we separate groups into two redshift bins, $z>0.5$ and $z\le 0.5$, where the total number of groups in these two redshift bins are roughly similar. We use red and blue histograms to represents results for groups within these two redshift bins, $z>0.5$ and $z\le 0.5$, respectively. We can see that the performance of $f_{\rm halo}$ depends 
only slightly on the redshift. 
The group finder successfully selects $\sim 70\%$ ($\sim 90\%$) of  the true halos  with  mass $\sim
10^{12}\msunh$ ($\ga
10^{14}\msunh$).  Note that this does not imply in any case that the $10 \sim 30\%$ of the remaining groups are not selected by our group finder but rather that their group membership completeness is lower than 50\%. If we lower the membership completeness criteria,  $f_{\rm halo}$ will increase.  
  
The second quantity we investigate is the purity of the groups, $f_{\rm group}$, defined as the fraction of groups having the 
richest components that contain more than half of the true members.  
{Here again, we separate groups into two redshift bins, $z>0.5$ and $z\le 0.5$, and use red and blue histograms to represents the related results.} 
The lower left panel of Fig.~\ref{fig:comp_halo} shows $f_{\rm
  group}$ obtained from our MGRS, for groups with $N_{\rm g} \geq 3, 5, 10$ as functions of the assigned halo mass for the case ${\rm B=5}$. The $f_{\rm
  group}$ does not show significant differences between the two redshift bins as well.  Overall, more than $90\%$ groups with mass $10^{12}\msunh$ are true groups. The percentage increases as the assigned halo mass and reaches 100\% for groups with mass larger than $10^{14.5}\msunh$. The remaining few groups are those split groups or those whose richest component have less than 50\% membership completeness. This value demonstrates that the detected groups regardless of their redshifts are in general reliable. 

Shown in the right-hand panels of Fig.~\ref{fig:comp_halo} are the corresponding $f_{\rm halo}$ and $f_{\rm group}$ for the case of ${\rm B=10}$. Overall, similar performance is achieved as those in the ${\rm B=5}$ case. Except the global completeness in the high redshift bin which is slightly lower, all the other quantities of the detected groups are similar. Note that since $f_{\rm halo}$  corresponds to the fraction of true groups being detected, while $f_{\rm group}$ corresponds to the fraction of the detected groups are true ones. In order to  taking into account the impact of incompleteness on the halo mass estimation using the abundance matching method, we correct the abundance of the groups by a factor of $f_{\rm incomp}=f_{\rm group}/f_{\rm halo}$. 


%
\begin{figure*}
\center
\includegraphics[height=13.0cm,width=13cm,angle=0]{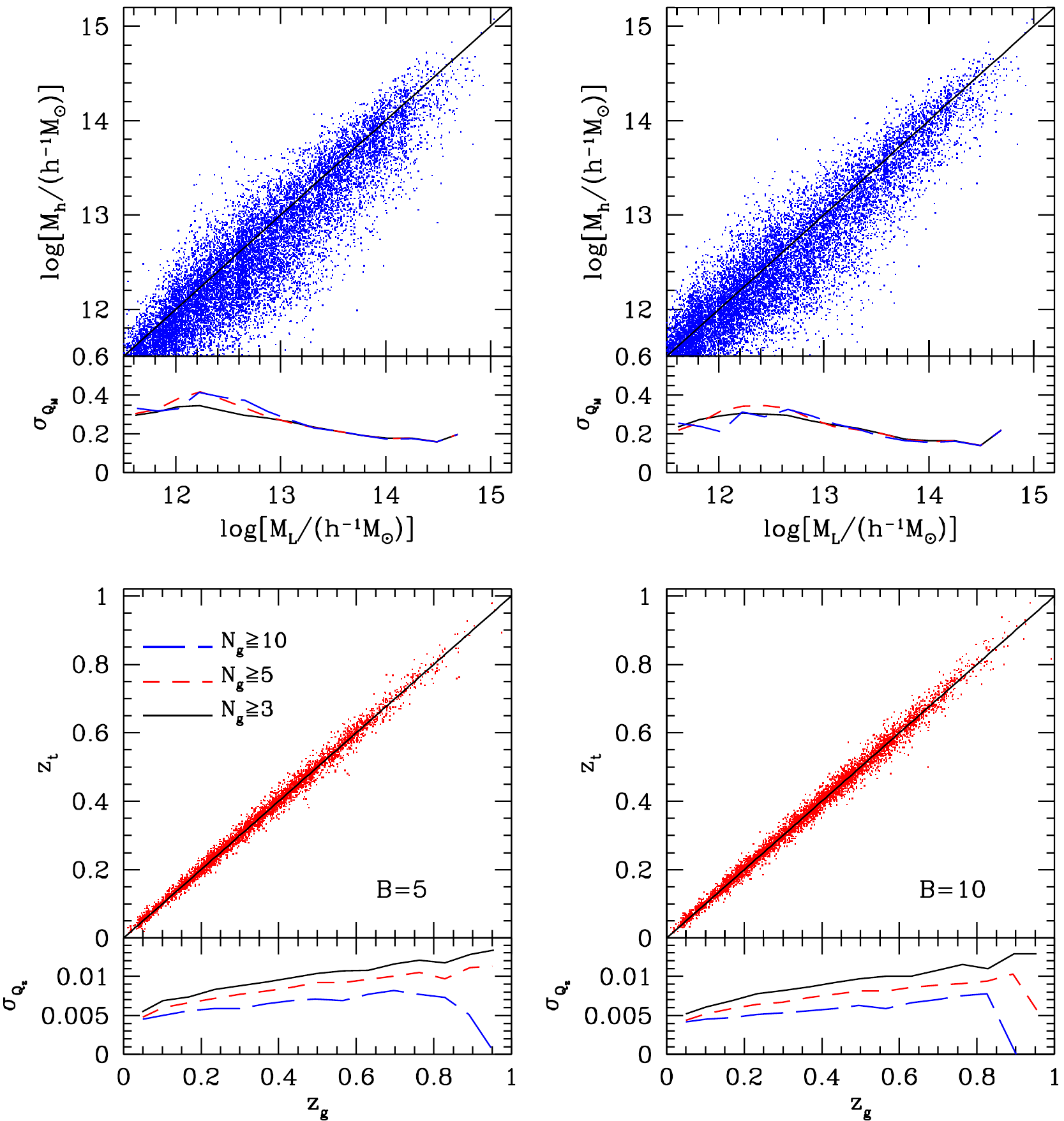}
\caption{Upper panels: Comparisons between the assigned halo mass ${\rm M_{L}}$ (based on the total group luminosity ${\rm L_{tot}}$)
  and the true halo mass ${\rm M_h}$. The dots are shown for a small subset of groups in the mock group catalog constructed from our MGRS. The small panels
  plot the scatter from the solid equality line as defined by equation~(\ref{QparamM}). Here, results are shown for groups with different richness cut limits as indicated. Lower panels: comparison between the assigned redshift of the group, ${\rm z_g}$, and the true halo redshift ${\rm z_t}$.  The small sub-panel describes the deviation from the solid equality line. The left and right  panels are the results obtained for background values ${\rm B=5}$ and ${\rm B=10}$,  respectively.}
\label{fig:M_true}
\end{figure*}

\subsection{The determination of halo mass and redshift}
\label{sec:mass&redshift}

In this subsection we investigate the reliability of the assigned halo mass and redshift determinations of the groups. First we check the assigned halo mass obtained in the group finding steps 2-3.
As pointed out in Y05, our halo mass estimation from the total group luminosity has the advantage 
that it is equally applicable to groups spanning a wide range in richness. However, a 
shortcoming of this method is that it requires the halo mass function, which is dependent on cosmology. In general, the assigned halo mass to each group can be regarded as the abundance of its population. Thus, it is extremely easy to convert the retrieved group mass in one cosmology to that in another cosmology, 
without running another round of group finder.

In order to assess the reliability of the halo mass assigned to individual groups, we use the mock group catalog from the MGRS. The top-left panel of Fig~\ref{fig:M_true} shows the assigned halo mass ${M_L}$  versus true halo mass ${M_h}$
in the ${\rm B=5}$ case. To quantitatively compare the scatter to the line of equality (${M_L  = M_h}$), we compute the following quantity for each group
\begin{equation}
\label{QparamM}
Q_M \equiv {1 \over \sqrt{2}} \left[\log(M_L) - \log(M_h) \right]
\end{equation}
and  measure the standard deviation, ${\rm \sigma_{Q_M}}$,  in several  bins of ${ \left[\log(M_L) +  \log(M_h) \right]/2}$. The results shown in the small panel indicate that the scatter is $\sim  0.30$ dex for groups with $10^{12} \msunh \la M_L  \la 10^{13.0} \msunh$, decreasing to $0.15$ dex at the high mass end. {Here we find that the $\sigma_{Q_M}$ is quite independent of the number of members in the groups,  according to the very similar behaviors shown for groups with different richness cuts. }

\begin{deluxetable*}{lcccccc}
\tabletypesize{\scriptsize}
\tablecaption{Number of Galaxies and Groups in the DESI image legacy surveys DR8 }
\tablewidth{0pt}
\tablehead{
\colhead{Region} &
\colhead{Galaxies} &
\colhead{Sky coverage} &
\colhead{Groups $N_g \ge 1$} &
\colhead{$N_g \ge 3$} &
\colhead{$N_g \ge 5$} &
\colhead{$N_g \ge 10$} \\
(1) & (2) & (3) & (4) & (5) & (6) & (7) 
}
\startdata
NGC	& 69753872 & 9673 &	53335347 &	2750934 &	879844 &	200674 \\
SGC & 59598763 & 8580 &	45022007 &	2421063 &	795323 &	186841
\enddata
\tablecomments{For each of the NGC or SGC samples, columns (2) and (3) 
  list the number and sky coverage (in unit of square degree) of galaxies.   
  In addition,  columns (4)--(7)  list the  numbers of
  groups with at least $1$, $3$, $5$, and $10$ members.}
\label{tab:number}
\end{deluxetable*}

There are several factors that contribute to this scatter. The first is the intrinsic scatter in the relation between the true halo mass and the true total group luminosity.
Here for simplicity, we did not correct for the faint galaxies incompleteness caused by the survey magnitude cut. As shown in the lower right panel of Fig.~\ref{fig:mgLF}, the mass-to-light ratio has a typical scatter  
of $0.15-0.2$ dex. Taking into account the $\sqrt{2}$ factor in Eq. \ref{QparamM}, this factor already contribute a lot to the scatter, especially at the high mass end.  

The second source of the scatter originates from the incompleteness and contamination introduced by our group finder, which may lead to biased total group luminosities.
Nevertheless, as we mentioned before, since we are using the halo abundance matching method to assign halo mass, the systematic difference will not impact our results. Only chaotic difference, 
e.g., fragmentation or mis-identification will impact the halo mass assignment significantly, which may occur at mass $\la 10^{13.0} \msunh$ \citep{Campbell2015}.

The last source of scatter in the assigned group masses owes to the fact that groups in the high redshift bin $0.67<z\le 1.0$ is incomplete (see the upper-left panel of Fig. \ref{fig:mgLF}). {Although we have used the $V_{\rm max}$ method to correct 
for the incompleteness of groups, since there are scatters among halo masses and BCG luminosities, such corrections are still not enough.  The assigned halo mass to our groups will be systematically higher, especially at the low mass end. }

Taking into all the above factors into account, we suggest that 
in our catalog, the groups with mass $\ga 10^{13.0} \msunh$ are more reliable. 

Next, we investigate the redshift accuracy of the groups. Shown in the lower left panel of Fig. \ref{fig:M_true} is the {scatter plot}
of true group redshift ${z_t}$ v.s. assigned group redshift ${z_g}$. {Here $z_g$ is defined as the luminosity weight group redshift.}
To quantify the scatter with respect to the  line of  equality  (${z_t=z_g}$), we follow a similar
track of halo mass scatter and compute the following quantity for each group
\begin{equation}
\label{eq:Qparamz}
Q_z \equiv {1 \over \sqrt{2}} {{z_t  - z_g} \over
{1+(z_t +  z_g)/2}}
\end{equation}
and  measure the standard  deviation, ${\sigma_{Q_z}}$,  in several  bins of ${\left[z_t +  z_g \right]/2}$. Comparing to
the input errors, the results shown in the small panel indicate  that the scatter drops with the
increase of member galaxies. The typical redshift error in groups with at least 10 members is
about 0.008  (taking into account the $\sqrt{2}$ factor in Eq. \ref{eq:Qparamz}). 

Finally, shown in the right hand panels of Fig. \ref{fig:M_true} are the corresponding mass and redshift  determinations for the case of ${\rm B=10}$. The general trends are quite similar to the ${\rm B=5}$ case, while the mass scatter is somewhat smaller, especially in relatively low mass groups with masses $\la 10^{13.0} \msunh$ at about 0.40 dex (taking into account the $\sqrt{2}$ factor in Eq. \ref{QparamM}). 

Combining all the above tests, we conclude that 
the ${\rm B=10}$ case performs better than the ${\rm B=5}$ case in the group membership purity and halo mass estimation, albeit with somewhat worse group membership completeness. Thus we adopt ${\rm B=10}$ value in our group finder for the Legacy Surveys DR8, while ${\rm B=5}$ will be used for future comparison studies.

\begin{figure*}
\center
\vspace{-0cm}
\includegraphics[height=15cm,width=15cm,angle=0]{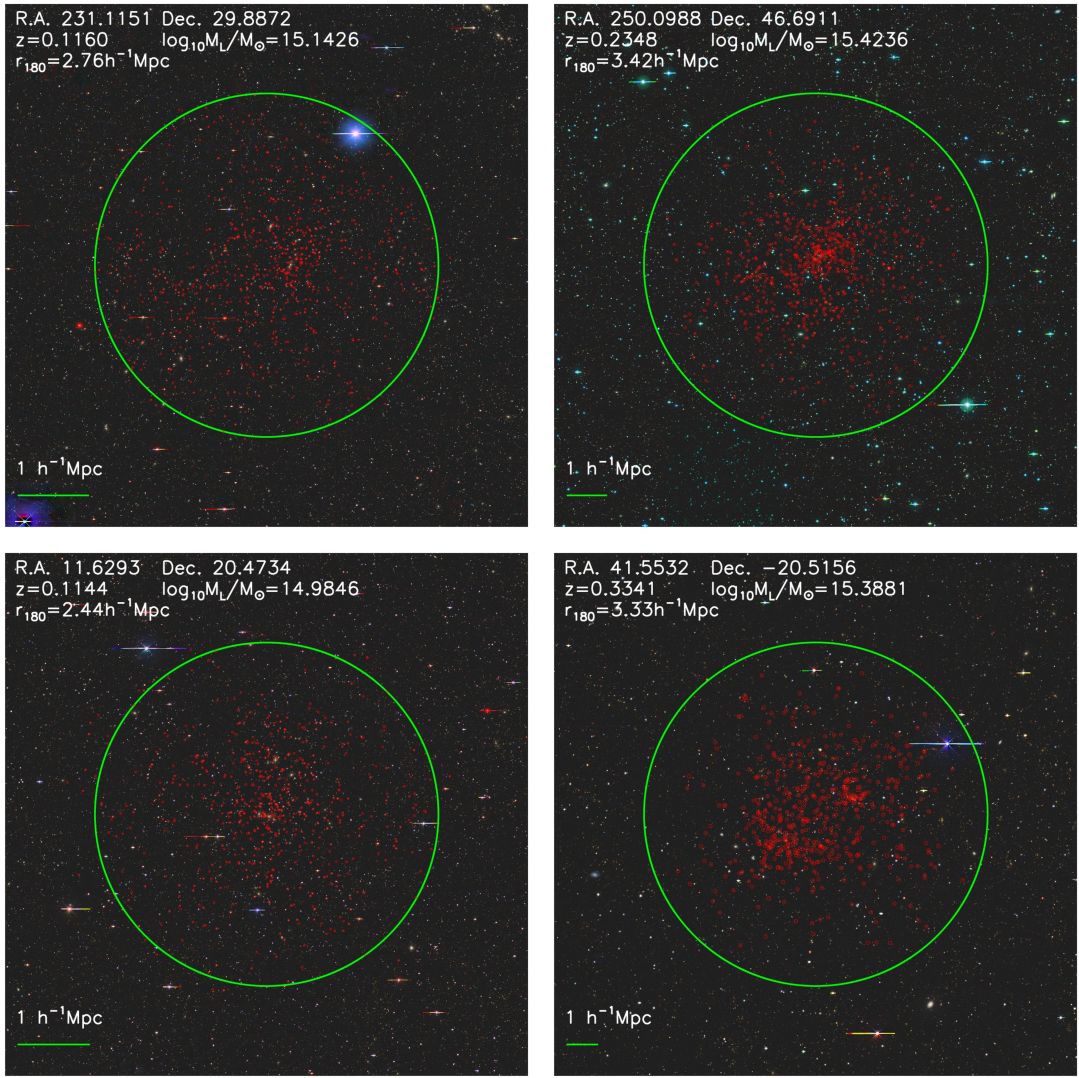}
\vspace{-0cm}
\caption{Image illustrations of four of the richest groups in our Legacy Surveys DR8 group catalogs. The upper and lower two panels are for two groups in the NGC and SGC, respectively. In each panel the background image is obtained from the Legacy Surveys DR8 image viewer. {The green circle illustrates the halo radius, $r_{180}$, of the group.} The small red circles mark the locations of the member galaxies. The location,  halo mass and redshift  information is marked on the upper-left corner 
of the image. }
\label{fig:cluster}
\end{figure*}

\section{Basic Properties of the Group Catalog}
\label{sec_catalogue}

We apply our extended halo-based group finder to the Legacy Surveys DR8 described in Section~\ref{sec:DESIDR8}.
In total, we obtain $53335347$ and 
$45022007$ groups in the NGC and SGC, respectively. In the following paragraphs we present some global properties of these group catalogues.

Table \ref{tab:number} lists the results for galaxy samples on the NGC and SGC, respectively for the number of groups with at least 1, 3, 5, and 10 members\footnote{Note that since we might update our group catalogs once additional spectroscopic redshifts
are available, the values listed in Tables 1 and 2 are subject to small changes.}. In total, there are about 5.2 Million groups  with at least 3 members, within which about 0.4 million  have at least 10 members.  

As an illustration, Fig.~\ref{fig:cluster}  shows the images of four richest groups in our catalog (two each on the NGC and SGC). In each 
panel the background image is obtained from the Legacy Surveys DR8 image viewer. The green circle on top of each image illustrates the {halo radius, $r_{180}$, of the group.} The location,  halo mass and redshift  information is marked on the upper-left corner of the image. As an example, the  richest group on NGC, located at ${\rm ra=231.1151}$,  ${\rm dec=29.8872}$ at redshift ${z=0.1160}$ has 697 member galaxies. For a better illustration, we also mark the locations of all the member galaxies with small red circles. With a careful and zoom in check of these clusters (use the Legacy Surveys DR8 image viewer directly), we can see that there are a few sub-clumps distributed within the halo radius indicating the existence of large sub-structures in massive halos.

\begin{figure*}
\center
\includegraphics[height=13.0cm,width=13cm,angle=0]{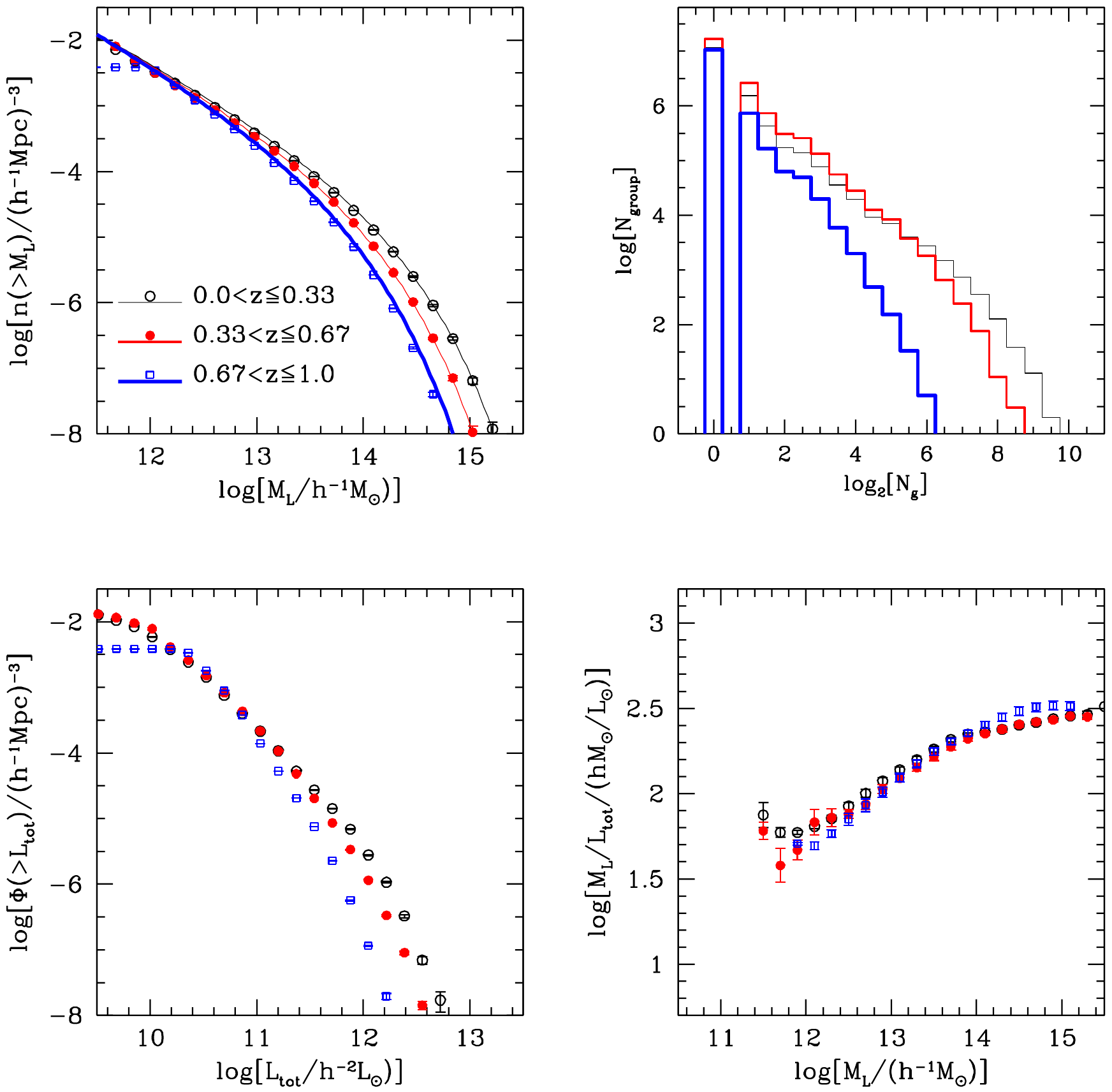}
\caption{Similar to Fig. \ref{fig:mgLF} but here the results are obtained from the Legacy Surveys DR8. The solid lines shown in the upper-left panel are obtained assuming a Planck18 cosmology.  }
\label{fig:DESILF}
\end{figure*} 

\begin{deluxetable*}{ccccccc}
\tabletypesize{\scriptsize} 
\tablecaption{The data structure of the Legacy Surveys DR8 group catalogs}

\tablewidth{0pt}
\tablehead{ 
Group ID & Richness & Ra & dec & redshift & Group mass & Group luminosity \\
(1) & (2) & (3) & (4) & (5) & (6) & (7)
}

\startdata
        1 & 697 & 231.1151 &  29.8872 &   0.1160  & 15.1426 & 500.3260\\
        2 & 630 & 250.0988 &  46.6911 &   0.2348  & 15.4236 & 818.2732\\
        3 & 586 & 229.1876 &  -0.8609 &   0.1216  & 15.1088 & 456.9565\\
        4 & 572 & 213.6720 &  -0.3643 &   0.1406  & 15.1395 & 496.9307\\
        5 & 554 & 127.6793 &  65.8586 &   0.1896  & 15.2778 & 652.1247\\
        6 & 486 & 203.8439 &  41.0002 &   0.2386  & 15.2548 & 631.4667\\
        7 & 486 & 249.0117 &  66.2045 &   0.1770  & 15.1457 & 503.9358\\
        8 & 477 & 260.6339 &  32.1059 &   0.2335  & 15.2154 & 589.6631\\
        ...\\
\cline{1-7}\\
        1 & 551 &  11.6293 &  20.4734  &  0.1144 &  14.9846 & 388.9367\\
        2 & 547 &  41.5532 & -20.5156  &  0.3341 &  15.3881 & 868.7377\\
        3 & 541 &  22.9696 & -13.5828  &  0.2186 &  15.4007 & 792.0868\\
        4 & 531 &   1.5695 & -34.7034  &  0.1214 &  14.9941 & 396.0752\\
        5 & 490 & 330.3961 & -59.9523  &  0.1066 &  14.9312 & 332.6521\\
        6 & 482 & 328.4296 & -57.7399  &  0.0791 &  14.8162 & 238.3979\\
        7 & 467 & 346.3691 & -44.2641  &  0.1373 &  14.9693 & 375.2200\\
        8 & 450 &   5.5730 &  23.2938  &  0.1388 &  14.9939 & 395.8793\\
        ...
\enddata

\tablecomments{Column (1): group ID.  Column (2): the richness of the group.  Columns (3)-(5): 
the luminosity weighted ra, dec and redshift of the group. Column (6): 
  logarithm of the assigned halo mass $\log M_L$. Columns (7): the total group 
  luminosity in unit of $10^{10}\Lsunhh$. Upper and lower parts are for groups
  in the NGC and SGC, respectively. }
  \label{tab:catalog}
\end{deluxetable*}

Next, we focus on a few statistical quantities for our groups. Shown in the upper-left panel of Fig.\ref{fig:DESILF} are the 
{\it accumulative} halo mass functions of groups at different redshift bins as indicated.  Following the MGRS case, we also show the 
corresponding theoretical model predictions by SMT2001 using solid lines for a Planck18 cosmology. The data agrees extremely well  with the model for the two low redshift bins and are somewhat underestimated at the high redshift bin. Note that since we are using the halo abundance matching method to assign the halo mass, they should agree by construction. The small deviation of the data from the theoretical model predictions for the high redshift bin originates from the {\it interpolation} of the mass-to-light ratios obtained from different redshift bins. 

Shown in the upper-right panel is the richness distribution of groups in the Legacy Surveys DR8. Here we can see that the richest group have about ${\rm N_{g}\sim 700}$ members in the low redshift bin, slightly less rich than that in the MGRS. Apart from a few richest clusters, these numbers are roughly consistent with those found in the MGRS after taking into the sky coverage into account. In the high redshift bin, the richest groups have around $30$ members which is quite similar to those in the MGRS. 

At the lower-left panel we present the {\it accumulative} total luminosity function of groups in the Legacy Surveys DR8. In the middle and high redshift bins, the groups with $L_{\rm tot}<10^{10.0}\Lsunhh$ and $L_{\rm tot}<10^{10.6}\Lsunhh$ are missing, which leads to a flattening feature at these luminosities.  

Shown at the lower-right panel are the mass-to-light ratio of groups in the Legacy Surveys DR8. The mass-to-light ratios in the two lower redshift bins in groups with mass $\ga 10^{13.5}\msunh$ are in good agreement with each other. While the one at the high redshift bin is higher by roughly $0.1$ dex. This is also the main reason that the halo mass function shown in the upper-left panel is underestimated. In groups with mass $\sim 10^{12.0}\msunh$, the mass-to-light ratios show upturns at the lower mass end, which is in general consistent with those CLF model predictions \citep[e.g.][]{Yang2003}. 

Finally, for any interested readers in our group catalogs, we have made them publicly available via the following link\footnote{https://gax.sjtu.edu.cn/data/data1/DESI\_DR8/groups.tar.gz}. The data structure and first lines in our catalog are described in
Table \ref{tab:catalog}.
Since our mass assignments require  the use of  the halo  mass function, which is cosmology  dependent, it is very easy to obtain the halo mass for other cosmologies by using the following relation:
\begin{equation}
\label{massconvert}
\int_{M_h}^{\infty} n(M'_h) \rmd M'_h = \int_{\widetilde{M_h}}^{\infty} 
\tilde{n}(M'_h) \rmd M'_h \,,
\end{equation}
where ${\rm M_h}$ and  ${\rm n(M_h)}$ are the mass and halo mass function in the Planck18 cosmology,   and   $\widetilde{M_h}$  and  $\tilde{n}(M_h)$  are the corresponding values in the other cosmology.
In addition to the main group catalogs, since we have spectroscopic redshifts for a fraction of galaxies, we will provide
a sub-catalog which contains the spectroscopic redshifts for the brightest member galaxies.

\section{Summary}
\label{sec_conclusion}

In this paper, we extended the halo-based group finder developed in Y05 and Y07, so that it can be applied to {galaxy samples with spectroscopic or photometric redshifts simultaneously.} The changes and/or improvements in our group finder are the following:
\begin{itemize}
\item We start with our group finder by assuming that all galaxies are tentative groups.  We use abundance matching between the total group luminosity and the halo mass to obtain the mass-to-light ratios in an iterative way.
\item In  order to minimize the impact of the survey magnitude limit to our halo mass estimation, we separate galaxy samples into different redshift bins and obtain their mass-to-light ratios separately. Interpolations in both redshift and group luminosity help obtain the halo mass for each group.
\item In the evaluation of the redshift distribution probability function, we also take into account the photometric redshift errors in addition to {the galaxy velocity dispersion in halos. This is indeed the most important extension in this version group finder, so that we can treat galaxies with spectroscopic and photometric redshifts equally. }
\item Apart from the fiducial background value (${\rm B=10}$), we also tested the case with a lower background value (${\rm B=5}$) to enhance the group membership completeness for the photometric redshift samples, which can be used for future comparisons.
\end{itemize}
{Our tests based on a 200 square degree mock galaxy redshift survey (MGRS) with a typical photometric redshift error of $(0.01+0.015z)*(1+z)$ show that for groups with mass $\ga 10^{12.5}\msunh$, $\sim 90\%$ of them have a membership completeness of $> 60\%$.  We
note that  $85\%$ of the groups with mass $\ga 10^{12.5}\msunh$ have an interloper (group members that belong to a different halo) fraction  $< 1$. In terms of group global properties the group finder can successfully detect more than half members in about $70\%$ and $80\%$ groups with mass  $\sim 10^{12.0}\msunh$ and $\ga 10^{14.0}\msunh$, respectively.} For the detected groups with masses $\ga 10^{12.0}\msunh$, the purity ($f_{\rm group}$) is larger than 90\%. The halo mass assigned to each group has an uncertainty of about 0.2 dex at the high mass end $\ga 10^{13.5}\msunh$ and increase to 0.40 dex at the low mass end. The redshift accuracy of the groups with at least 10 members is at $\sim 0.008$.

We applied our extended halo-based group finder to the DESI Legacy Imaging Surveys DR8. We obtain separate group catalogs on the north galactic cap (NGC) and south galactic cap (SGC) samples. In the group catalogs, each group is assigned a halo mass, average ra, dec and redshift, etc. Based on these group catalogs, we investigate their basic properties, such as the distributions of richness, halo mass, and luminosity. In addition, we have presented the average ratios between halo mass  and group luminosity. A detailed analysis of the group properties and their implications related to halo occupation statistics, galaxy formation and evolution, and cosmology will be presented in a series of forthcoming papers.


\section*{Acknowledgments}

We thank the anonymous referee for helpful comments that greatly improved the presentation of this paper.
This work is supported by the national science foundation of China (Nos. 11833005,  11890691, 11890692, 11621303, 11890693, 11421303), 111 project No. B20019 and Shanghai Natural Science Foundation, grant No. 15ZR1446700, 19ZR1466800. This work is also supported by the U.D Department of Energy, Office of Science, Office of High Energy Physics under Award Number DE-SC0019301.

This work made use of the Gravity Supercomputer at the Department of Astronomy, Shanghai Jiao Tong University.

The Photometric Redshifts for the Legacy Surveys (PRLS) catalog used in this paper was produced thanks to funding from the U.S. Department of Energy Office of Science, Office of High Energy Physics via grant DE-SC0007914.


\bibliography{Mybibtex.bib}

\begin{thebibliography}{}
\expandafter\ifx\csname natexlab\endcsname\relax\def\natexlab#1{#1}\fi
\providecommand{\url}[1]{\href{#1}{#1}}
\providecommand{\dodoi}[1]{doi:~\href{http://doi.org/#1}{\nolinkurl{#1}}}
\providecommand{\doeprint}[1]{\href{http://ascl.net/#1}{\nolinkurl{http://ascl.net/#1}}}
\providecommand{\doarXiv}[1]{\href{https://arxiv.org/abs/#1}{\nolinkurl{https://arxiv.org/abs/#1}}}

\bibitem[{{Avila-Reese} {et~al.}(2011){Avila-Reese}, {Col{\'\i}n},
  {Gonz{\'a}lez-Samaniego}, {Valenzuela}, {Firmani}, {Vel{\'a}zquez}, \&
  {Ceverino}}]{Avila-Reese2011}
{Avila-Reese}, V., {Col{\'\i}n}, P., {Gonz{\'a}lez-Samaniego}, A., {et~al.}
  2011, \apj, 736, 134, \dodoi{10.1088/0004-637X/736/2/134}

\bibitem[{{Bahcall} {et~al.}(2003){Bahcall}, {McKay}, {Annis}, {Kim}, {Dong},
  {Hansen}, {Goto}, {Gunn}, {Miller}, {Nichol}, {Postman}, {Schneider},
  {Schroeder}, {Voges}, {Brinkmann}, \& {Fukugita}}]{Bahcall2003}
{Bahcall}, N.~A., {McKay}, T.~A., {Annis}, J., {et~al.} 2003, \apjs, 148, 243,
  \dodoi{10.1086/377167}

\bibitem[{{Berlind} {et~al.}(2006){Berlind}, {Frieman}, {Weinberg}, {Blanton},
  {Warren}, {Abazajian}, {Scranton}, {Hogg}, {Scoccimarro}, {Bahcall},
  {Brinkmann}, {Gott}, {Kleinman}, {Krzesinski}, {Lee}, {Miller}, {Nitta},
  {Schneider}, {Tucker}, {Zehavi}, \& {SDSS Collaboration}}]{Berlind2006}
{Berlind}, A.~A., {Frieman}, J., {Weinberg}, D.~H., {et~al.} 2006, \apjs, 167,
  1, \dodoi{10.1086/508170}

\bibitem[{{Blanton} \& {Roweis}(2007)}]{Blanton2007}
{Blanton}, M.~R., \& {Roweis}, S. 2007, \aj, 133, 734, \dodoi{10.1086/510127}

\bibitem[{{Brown} {et~al.}(2008){Brown}, {Zheng}, {White}, {Dey}, {Jannuzi},
  {Benson}, {Brand }, {Brodwin}, \& {Croton}}]{Brown2008}
{Brown}, M. J.~I., {Zheng}, Z., {White}, M., {et~al.} 2008, \apj, 682, 937,
  \dodoi{10.1086/589538}

\bibitem[{{Cacciato} {et~al.}(2009){Cacciato}, {van den Bosch}, {More}, {Li},
  {Mo}, \& {Yang}}]{Cacciato2009}
{Cacciato}, M., {van den Bosch}, F.~C., {More}, S., {et~al.} 2009, \mnras, 394,
  929, \dodoi{10.1111/j.1365-2966.2008.14362.x}

\bibitem[{Campbell {et~al.}(2015)Campbell, van~den Bosch, Hearin, Padmanabhan,
  Berlind, Mo, Tinker, \& Yang}]{Campbell2015}
Campbell, D., van~den Bosch, F.~C., Hearin, A., {et~al.} 2015, Monthly Notices
  of the Royal Astronomical Society, 452, 444

\bibitem[{{Chen} {et~al.}(2019){Chen}, {Mo}, {Li}, {Wang}, {Yang}, {Zhou}, \&
  {Zhang}}]{Chen2019elu}
{Chen}, Y., {Mo}, H.~J., {Li}, C., {et~al.} 2019, \apj, 872, 180,
  \dodoi{10.3847/1538-4357/ab0208}

\bibitem[{{Coil} {et~al.}(2006){Coil}, {Gerke}, {Newman}, {Ma}, {Yan},
  {Cooper}, {Davis}, {Faber}, {Guhathakurta}, \& {Koo}}]{Coil2006}
{Coil}, A.~L., {Gerke}, B.~F., {Newman}, J.~A., {et~al.} 2006, \apj, 638, 668,
  \dodoi{10.1086/498885}

\bibitem[{{Colless} {et~al.}(2001){Colless}, {Dalton}, {Maddox}, {Sutherland },
  {Norberg}, {Cole}, {Bland -Hawthorn}, {Bridges}, {Cannon}, {Collins},
  {Couch}, {Cross}, {Deeley}, {De Propris}, {Driver}, {Efstathiou}, {Ellis},
  {Frenk}, {Glazebrook}, {Jackson}, {Lahav}, {Lewis}, {Lumsden}, {Madgwick},
  {Peacock}, {Peterson}, {Price}, {Seaborne}, \& {Taylor}}]{Colless2001}
{Colless}, M., {Dalton}, G., {Maddox}, S., {et~al.} 2001, \mnras, 328, 1039,
  \dodoi{10.1046/j.1365-8711.2001.04902.x}

\bibitem[{{Collister} \& {Lahav}(2005)}]{Collister2005}
{Collister}, A.~A., \& {Lahav}, O. 2005, \mnras, 361, 415,
  \dodoi{10.1111/j.1365-2966.2005.09172.x}

\bibitem[{{Crook} {et~al.}(2007){Crook}, {Huchra}, {Martimbeau}, {Masters},
  {Jarrett}, \& {Macri}}]{Crook2007}
{Crook}, A.~C., {Huchra}, J.~P., {Martimbeau}, N., {et~al.} 2007, \apj, 655,
  790, \dodoi{10.1086/510201}

\bibitem[{Darvish {et~al.}(2017)Darvish, Mobasher, Martin, Sobral, Scoville,
  Stroe, Hemmati, \& Kartaltepe}]{Darvish2017}
Darvish, B., Mobasher, B., Martin, D.~C., {et~al.} 2017, The Astrophysical
  Journal, 837, 16, \dodoi{10.3847/1538-4357/837/1/16}

\bibitem[{{Davis} {et~al.}(1985){Davis}, {Efstathiou}, {Frenk}, \&
  {White}}]{Davis1985}
{Davis}, M., {Efstathiou}, G., {Frenk}, C.~S., \& {White}, S.~D.~M. 1985, \apj,
  292, 371, \dodoi{10.1086/163168}

\bibitem[{{Dey} {et~al.}(2019){Dey}, {Schlegel}, {Lang}, {Blum}, {Burleigh},
  {Fan}, {Findlay}, {Finkbeiner}, {Herrera}, {Juneau}, {Landriau}, {Levi},
  {McGreer}, {Meisner}, {Myers}, {Moustakas}, {Nugent}, {Patej}, {Schlafly},
  {Walker}, {Valdes}, {Weaver}, {Y{\`e}che}, {Zou}, {Zhou}, {Abareshi},
  {Abbott}, {Abolfathi}, {Aguilera}, {Alam}, {Allen}, {Alvarez}, {Annis},
  {Ansarinejad}, {Aubert}, {Beechert}, {Bell}, {BenZvi}, {Beutler}, {Bielby},
  {Bolton}, {Brice{\~n}o}, {Buckley-Geer}, {Butler}, {Calamida}, {Carlberg},
  {Carter}, {Casas}, {Castander}, {Choi}, {Comparat}, {Cukanovaite}, {Delubac},
  {DeVries}, {Dey}, {Dhungana}, {Dickinson}, {Ding}, {Donaldson}, {Duan},
  {Duckworth}, {Eftekharzadeh}, {Eisenstein}, {Etourneau}, {Fagrelius},
  {Farihi}, {Fitzpatrick}, {Font-Ribera}, {Fulmer}, {G{\"a}nsicke},
  {Gaztanaga}, {George}, {Gerdes}, {Gontcho}, {Gorgoni}, {Green}, {Guy},
  {Harmer}, {Hernand ez}, {Honscheid}, {Huang}, {James}, {Jannuzi}, {Jiang},
  {Joyce}, {Karcher}, {Karkar}, {Kehoe}, {Kneib}, {Kueter-Young}, {Lan},
  {Lauer}, {Le Guillou}, {Le Van Suu}, {Lee}, {Lesser}, {Perreault Levasseur},
  {Li}, {Mann}, {Marshall}, {Mart{\'\i}nez-V{\'a}zquez}, {Martini}, {du Mas des
  Bourboux}, {McManus}, {Meier}, {M{\'e}nard}, {Metcalfe},
  {Mu{\~n}oz-Guti{\'e}rrez}, {Najita}, {Napier}, {Narayan}, {Newman}, {Nie},
  {Nord}, {Norman}, {Olsen}, {Paat}, {Palanque-Delabrouille}, {Peng},
  {Poppett}, {Poremba}, {Prakash}, {Rabinowitz}, {Raichoor}, {Rezaie},
  {Robertson}, {Roe}, {Ross}, {Ross}, {Rudnick}, {Safonova}, {Saha},
  {S{\'a}nchez}, {Savary}, {Schweiker}, {Scott}, {Seo}, {Shan}, {Silva},
  {Slepian}, {Soto}, {Sprayberry}, {Staten}, {Stillman}, {Stupak}, {Summers},
  {Sien Tie}, {Tirado}, {Vargas-Maga{\~n}a}, {Vivas}, {Wechsler}, {Williams},
  {Yang}, {Yang}, {Yapici}, {Zaritsky}, {Zenteno}, {Zhang}, {Zhang}, {Zhou}, \&
  {Zhou}}]{Dey2019}
{Dey}, A., {Schlegel}, D.~J., {Lang}, D., {et~al.} 2019, \aj, 157, 168,
  \dodoi{10.3847/1538-3881/ab089d}

\bibitem[{{D{\'\i}az-Gim{\'e}nez} \& {Zandivarez}(2015)}]{Diaz-Gimenez2015}
{D{\'\i}az-Gim{\'e}nez}, E., \& {Zandivarez}, A. 2015, \aap, 578, A61,
  \dodoi{10.1051/0004-6361/201425267}

\bibitem[{Duarte \& Mamon(2015)}]{Duarte2015}
Duarte, M., \& Mamon, G.~A. 2015, Monthly Notices of the Royal Astronomical
  Society, 453, 3849

\bibitem[{{Einasto} {et~al.}(2007){Einasto}, {Einasto}, {Tago}, {Saar},
  {H{\"u}tsi}, {J{\~o}eveer}, {Liivam{\"a}gi}, {Suhhonenko}, {Jaaniste},
  {Hein{\"a}m{\"a}ki}, {M{\"u}ller}, {Knebe}, \& {Tucker}}]{Einasto2007}
{Einasto}, J., {Einasto}, M., {Tago}, E., {et~al.} 2007, \aap, 462, 811,
  \dodoi{10.1051/0004-6361:20065296}

\bibitem[{{Eke} {et~al.}(2004){Eke}, {Baugh}, {Cole}, {Frenk}, {Norberg},
  {Peacock}, {Baldry}, {Bland-Hawthorn}, {Bridges}, {Cannon}, {Colless},
  {Collins}, {Couch}, {Dalton}, {de Propris}, {Driver}, {Efstathiou}, {Ellis},
  {Glazebrook}, {Jackson}, {Lahav}, {Lewis}, {Lumsden}, {Maddox}, {Madgwick},
  {Peterson}, {Sutherland}, \& {Taylor}}]{Eke2004}
{Eke}, V.~R., {Baugh}, C.~M., {Cole}, S., {et~al.} 2004, \mnras, 348, 866,
  \dodoi{10.1111/j.1365-2966.2004.07408.x}

\bibitem[{{Geller} \& {Huchra}(1983)}]{Geller1983}
{Geller}, M.~J., \& {Huchra}, J.~P. 1983, \apjs, 52, 61, \dodoi{10.1086/190859}

\bibitem[{{Gerke} {et~al.}(2005){Gerke}, {Newman}, {Davis}, {Marinoni}, {Yan},
  {Coil}, {Conroy}, {Cooper}, {Faber}, {Finkbeiner}, {Guhathakurta}, {Kaiser},
  {Koo}, {Phillips}, {Weiner}, \& {Willmer}}]{Gerke2005}
{Gerke}, B.~F., {Newman}, J.~A., {Davis}, M., {et~al.} 2005, \apj, 625, 6,
  \dodoi{10.1086/429579}

\bibitem[{{Goto}(2005)}]{Goto2005}
{Goto}, T. 2005, \mnras, 359, 1415, \dodoi{10.1111/j.1365-2966.2005.08982.x}

\bibitem[{{Goto} {et~al.}(2002){Goto}, {Sekiguchi}, {Nichol}, {Bahcall}, {Kim},
  {Annis}, {Ivezi{\'c}}, {Brinkmann}, {Hennessy}, {Szokoly}, \&
  {Tucker}}]{Goto2002}
{Goto}, T., {Sekiguchi}, M., {Nichol}, R.~C., {et~al.} 2002, \aj, 123, 1807,
  \dodoi{10.1086/339303}

\bibitem[{Han {et~al.}(2015)Han, Eke, Frenk, Mandelbaum, Norberg, Schneider,
  Peacock, Jing, Baldry, Bland-Hawthorn, Brough, Brown, Liske, Loveday, \&
  Robotham}]{Han2015}
Han, J., Eke, V.~R., Frenk, C.~S., {et~al.} 2015, Monthly Notices of the Royal
  Astronomical Society, 446, 1356, \dodoi{10.1093/mnras/stu2178}

\bibitem[{{Hao} {et~al.}(2010){Hao}, {McKay}, {Koester}, {Rykoff}, {Rozo},
  {Annis}, {Wechsler}, {Evrard}, {Siegel}, {Becker}, {Busha}, {Gerdes},
  {Johnston}, \& {Sheldon}}]{Hao2010}
{Hao}, J., {McKay}, T.~A., {Koester}, B.~P., {et~al.} 2010, \apjs, 191, 254,
  \dodoi{10.1088/0067-0049/191/2/254}

\bibitem[{{Hearin} \& {Watson}(2013)}]{Hearin2013}
{Hearin}, A.~P., \& {Watson}, D.~F. 2013, \mnras, 435, 1313,
  \dodoi{10.1093/mnras/stt1374}

\bibitem[{{Huchra} {et~al.}(2012){Huchra}, {Macri}, {Masters}, {Jarrett},
  {Berlind}, {Calkins}, {Crook}, {Cutri}, {Erdo{\v{g}}du}, {Falco}, {George},
  {Hutcheson}, {Lahav}, {Mader}, {Mink}, {Martimbeau}, {Schneider},
  {Skrutskie}, {Tokarz}, \& {Westover}}]{Huchra2012}
{Huchra}, J.~P., {Macri}, L.~M., {Masters}, K.~L., {et~al.} 2012, \apjs, 199,
  26, \dodoi{10.1088/0067-0049/199/2/26}

\bibitem[{{Jing} {et~al.}(1998){Jing}, {Mo}, \& {B{\"o}rner}}]{Jing1998}
{Jing}, Y.~P., {Mo}, H.~J., \& {B{\"o}rner}, G. 1998, \apj, 494, 1,
  \dodoi{10.1086/305209}

\bibitem[{{Jones} {et~al.}(2009){Jones}, {Read}, {Saunders}, {Colless},
  {Jarrett}, {Parker}, {Fairall}, {Mauch}, {Sadler}, {Watson}, {Burton},
  {Campbell}, {Cass}, {Croom}, {Dawe}, {Fiegert}, {Frankcombe}, {Hartley},
  {Huchra}, {James}, {Kirby}, {Lahav}, {Lucey}, {Mamon}, {Moore}, {Peterson},
  {Prior}, {Proust}, {Russell}, {Safouris}, {Wakamatsu}, {Westra}, \&
  {Williams}}]{Jones2009}
{Jones}, D.~H., {Read}, M.~A., {Saunders}, W., {et~al.} 2009, \mnras, 399, 683,
  \dodoi{10.1111/j.1365-2966.2009.15338.x}

\bibitem[{{Katsianis} {et~al.}(2020){Katsianis}, {Xu}, {Yang}, {Luo}, {Cui},
  {Dav{\'e}}, {Lagos}, {Zheng}, \& {Zhao}}]{Katsianis2020b}
{Katsianis}, A., {Xu}, H., {Yang}, X., {et~al.} 2020, arXiv e-prints,
  arXiv:2010.08173.
\newblock \doarXiv{2010.08173}

\bibitem[{Kawinwanichakij {et~al.}(2016)Kawinwanichakij, Quadri, Papovich,
  Kacprzak, Labbé, Spitler, Straatman, Tran, Allen, Behroozi, Cowley, Dekel,
  Glazebrook, Hartley, Kelson, Koo, Lee, Lu, Nanayakkara, Persson, Primack,
  Tilvi, Tomczak, \& van Dokkum}]{Kawinwanichakij2016}
Kawinwanichakij, L., Quadri, R.~F., Papovich, C., {et~al.} 2016, The
  Astrophysical Journal, 817, 9, \dodoi{10.3847/0004-637X/817/1/9}

\bibitem[{{Kim} {et~al.}(2002){Kim}, {Kepner}, {Postman}, {Strauss}, {Bahcall},
  {Gunn}, {Lupton}, {Annis}, {Nichol}, {Castander}, {Brinkmann}, {Brunner},
  {Connolly}, {Csabai}, {Hindsley}, {Ivezi{\'c}}, {Vogeley}, \&
  {York}}]{Kim2002}
{Kim}, R. S.~J., {Kepner}, J.~V., {Postman}, M., {et~al.} 2002, \aj, 123, 20,
  \dodoi{10.1086/324727}

\bibitem[{Knobel {et~al.}(2015)Knobel, Lilly, Woo, \& Kovač}]{Knobel2015}
Knobel, C., Lilly, S.~J., Woo, J., \& Kovač, K. 2015, The Astrophysical
  Journal, 800, 24, \dodoi{10.1088/0004-637X/800/1/24}

\bibitem[{Knobel {et~al.}(2012)Knobel, Lilly, Iovino, Kova{\v c}, Bschorr,
  Presotto, Oesch, Kampczyk, Carollo, Contini, Kneib, Le~F{\`e}vre, Mainieri,
  Renzini, Scodeggio, Zamorani, Bardelli, Bolzonella, Bongiorno, Caputi,
  Cucciati, de~la Torre, de~Ravel, Franzetti, Garilli, Lamareille, Le~Borgne,
  Le~Brun, Maier, Mignoli, Pell{\`o}, Peng, Perez-Montero, Silverman, Tanaka,
  Tasca, Tresse, Vergani, Zucca, Barnes, Bordoloi, Cappi, Cimatti, Coppa,
  Koekemoer, L{\'o}pez-Sanjuan, McCracken, Moresco, Nair, Pozzetti, \&
  Welikala}]{Knobel2012}
Knobel, C., Lilly, S.~J., Iovino, A., {et~al.} 2012, The Astrophysical Journal,
  753, 121

\bibitem[{{Koester} {et~al.}(2007){Koester}, {McKay}, {Annis}, {Wechsler},
  {Evrard}, {Bleem}, {Becker}, {Johnston}, {Sheldon}, {Nichol}, {Miller},
  {Scranton}, {Bahcall}, {Barentine}, {Brewington}, {Brinkmann}, {Harvanek},
  {Kleinman}, {Krzesinski}, {Long}, {Nitta}, {Schneider}, {Sneddin}, {Voges},
  \& {York}}]{Koester2007}
{Koester}, B.~P., {McKay}, T.~A., {Annis}, J., {et~al.} 2007, \apj, 660, 239,
  \dodoi{10.1086/509599}

\bibitem[{{Lacey} \& {Cole}(1993)}]{Lacey1993}
{Lacey}, C., \& {Cole}, S. 1993, \mnras, 262, 627,
  \dodoi{10.1093/mnras/262.3.627}

\bibitem[{Laigle {et~al.}(2016)Laigle, McCracken, Ilbert, Hsieh, Davidzon,
  Capak, Hasinger, Silverman, Pichon, Coupon, Aussel, Le~Borgne, Caputi,
  Cassata, Chang, Civano, Dunlop, Fynbo, Kartaltepe, Koekemoer, Le~F{\`e}vre,
  Le~Floc'h, Leauthaud, Lilly, Lin, Marchesi, {Milvang-Jensen}, Salvato,
  Sanders, Scoville, Smolcic, Stockmann, Taniguchi, Tasca, Toft, Vaccari, \&
  Zabl}]{Laigle2016}
Laigle, C., McCracken, H.~J., Ilbert, O., {et~al.} 2016, The Astrophysical
  Journal Supplement Series, 224, 24, \dodoi{10.3847/0067-0049/224/2/24}

\bibitem[{Lan {et~al.}(2016)Lan, M{\'e}nard, \& Mo}]{Lan2016}
Lan, T.-W., M{\'e}nard, B., \& Mo, H. 2016, Monthly Notices of the Royal
  Astronomical Society, 459, 3998

\bibitem[{{Lang} {et~al.}(2016){Lang}, {Hogg}, \& {Mykytyn}}]{Lang2016}
{Lang}, D., {Hogg}, D.~W., \& {Mykytyn}, D. 2016, {The Tractor: Probabilistic
  astronomical source detection and measurement}.
\newblock \doeprint{1604.008}

\bibitem[{{Leauthaud} {et~al.}(2012){Leauthaud}, {Tinker}, {Bundy}, {Behroozi},
  {Massey}, {Rhodes}, {George}, {Kneib}, {Benson}, {Wechsler}, {Busha},
  {Capak}, {Cort{\^e}s}, {Ilbert}, {Koekemoer}, {Le F{\`e}vre}, {Lilly},
  {McCracken}, {Salvato}, {Schrabback}, {Scoville}, {Smith}, \&
  {Taylor}}]{Leauthaud2012}
{Leauthaud}, A., {Tinker}, J., {Bundy}, K., {et~al.} 2012, \apj, 744, 159,
  \dodoi{10.1088/0004-637X/744/2/159}

\bibitem[{{Lee} {et~al.}(2004){Lee}, {Allam}, {Tucker}, {Annis}, {Johnston},
  {Scranton}, {Acebo}, {Bahcall}, {Bartelmann}, {B{\"o}hringer}, {Ellman},
  {Grebel}, {Infante}, {Loveday}, {McKay}, {Prada}, {Schneider}, {Stoughton},
  {Szalay}, {Vogeley}, {Voges}, \& {Yanny}}]{Lee2004}
{Lee}, B.~C., {Allam}, S.~S., {Tucker}, D.~L., {et~al.} 2004, \aj, 127, 1811,
  \dodoi{10.1086/382236}

\bibitem[{Li {et~al.}(2011)Li, Mo, Fan, Bosch, \& Yang}]{Li2011}
Li, R., Mo, H.~J., Fan, Z., Bosch, F. C. v.~d., \& Yang, X. 2011, Monthly
  Notices of the Royal Astronomical Society, 413, 3039,
  \dodoi{10.1111/j.1365-2966.2011.18378.x}

\bibitem[{Lim {et~al.}(2020)Lim, Mo, Wang, \& Yang}]{Lim2020}
Lim, S., Mo, H., Wang, H., \& Yang, X. 2020, The Astrophysical Journal, 889, 48

\bibitem[{Lim {et~al.}(2018)Lim, Mo, Li, Liu, Ma, Wang, \& Yang}]{Lim2018}
Lim, S.~H., Mo, H.~J., Li, R., {et~al.} 2018, The Astrophysical Journal, 854,
  181, \dodoi{10.3847/1538-4357/aaaa21}

\bibitem[{{Lim} {et~al.}(2017){Lim}, {Mo}, {Lu}, {Wang}, \& {Yang}}]{Lim2017}
{Lim}, S.~H., {Mo}, H.~J., {Lu}, Y., {Wang}, H., \& {Yang}, X. 2017, \mnras,
  470, 2982, \dodoi{10.1093/mnras/stx1462}

\bibitem[{{Lu} {et~al.}(2015){Lu}, {Yang}, \& {Shen}}]{Lu2015}
{Lu}, Y., {Yang}, X., \& {Shen}, S. 2015, \apj, 804, 55,
  \dodoi{10.1088/0004-637X/804/1/55}

\bibitem[{{Lu} {et~al.}(2016){Lu}, {Yang}, {Shi}, {Mo}, {Tweed}, {Wang},
  {Zhang}, {Li}, \& {Lim}}]{Lu2016}
{Lu}, Y., {Yang}, X., {Shi}, F., {et~al.} 2016, \apj, 832, 39,
  \dodoi{10.3847/0004-637X/832/1/39}

\bibitem[{Luo {et~al.}(2018)Luo, Yang, Lu, Shi, Zhang, Mo, Shu, Fu, Radovich,
  Zhang, {et~al.}}]{Luo2018}
Luo, W., Yang, X., Lu, T., {et~al.} 2018, The Astrophysical Journal, 862, 4

\bibitem[{{Macci{\`o}} {et~al.}(2007){Macci{\`o}}, {Dutton}, {van den Bosch},
  {Moore}, {Potter}, \& {Stadel}}]{Maccio2007}
{Macci{\`o}}, A.~V., {Dutton}, A.~A., {van den Bosch}, F.~C., {et~al.} 2007,
  \mnras, 378, 55, \dodoi{10.1111/j.1365-2966.2007.11720.x}

\bibitem[{Mandelbaum {et~al.}(2006)Mandelbaum, Seljak, Cool, Blanton, Hirata,
  \& Brinkmann}]{Mandelbaum2006}
Mandelbaum, R., Seljak, U., Cool, R.~J., {et~al.} 2006, Monthly Notices of the
  Royal Astronomical Society, 372, 758,
  \dodoi{10.1111/j.1365-2966.2006.10906.x}

\bibitem[{{Meng} {et~al.}(2020){Meng}, {Li}, {Mo}, {Chen}, \&
  {Wang}}]{Meng2020}
{Meng}, J., {Li}, C., {Mo}, H., {Chen}, Y., \& {Wang}, K. 2020, arXiv e-prints,
  arXiv:2008.13733.
\newblock \doarXiv{2008.13733}

\bibitem[{{Merch{\'a}n} \& {Zandivarez}(2002)}]{Merchan2002}
{Merch{\'a}n}, M., \& {Zandivarez}, A. 2002, \mnras, 335, 216,
  \dodoi{10.1046/j.1365-8711.2002.05623.x}

\bibitem[{{Merch{\'a}n} \& {Zandivarez}(2005)}]{Merchan2005}
{Merch{\'a}n}, M.~E., \& {Zandivarez}, A. 2005, \apj, 630, 759,
  \dodoi{10.1086/427989}

\bibitem[{{Miller} {et~al.}(2005){Miller}, {Nichol}, {Reichart}, {Wechsler},
  {Evrard}, {Annis}, {McKay}, {Bahcall}, {Bernardi}, {Boehringer}, {Connolly},
  {Goto}, {Kniazev}, {Lamb}, {Postman}, {Schneider}, {Sheth}, \&
  {Voges}}]{Miller2005}
{Miller}, C.~J., {Nichol}, R.~C., {Reichart}, D., {et~al.} 2005, \aj, 130, 968,
  \dodoi{10.1086/431357}

\bibitem[{{Mo} {et~al.}(2010){Mo}, {van den Bosch}, \& {White}}]{Mo2010}
{Mo}, H., {van den Bosch}, F.~C., \& {White}, S. 2010, {Galaxy Formation and
  Evolution}

\bibitem[{{More} {et~al.}(2009){More}, {van den Bosch}, {Cacciato}, {Mo},
  {Yang}, \& {Li}}]{More2009}
{More}, S., {van den Bosch}, F.~C., {Cacciato}, M., {et~al.} 2009, \mnras, 392,
  801, \dodoi{10.1111/j.1365-2966.2008.14095.x}

\bibitem[{{Navarro} {et~al.}(1997){Navarro}, {Frenk}, \& {White}}]{Navarro1997}
{Navarro}, J.~F., {Frenk}, C.~S., \& {White}, S. D.~M. 1997, \apj, 490, 493,
  \dodoi{10.1086/304888}

\bibitem[{{Neistein} {et~al.}(2011){Neistein}, {Weinmann}, {Li}, \&
  {Boylan-Kolchin}}]{Neistein2011}
{Neistein}, E., {Weinmann}, S.~M., {Li}, C., \& {Boylan-Kolchin}, M. 2011,
  \mnras, 414, 1405, \dodoi{10.1111/j.1365-2966.2011.18473.x}

\bibitem[{{Oguri} {et~al.}(2018){Oguri}, {Lin}, {Lin}, {Nishizawa}, {More},
  {More}, {Hsieh}, {Medezinski}, {Miyatake}, {Jian}, {Lin}, {Takada}, {Okabe},
  {Speagle}, {Coupon}, {Leauthaud}, {Lupton}, {Miyazaki}, {Price}, {Tanaka},
  {Chiu}, {Komiyama}, {Okura}, {Tanaka}, \& {Usuda}}]{Oguri2018}
{Oguri}, M., {Lin}, Y.-T., {Lin}, S.-C., {et~al.} 2018, \pasj, 70, S20,
  \dodoi{10.1093/pasj/psx042}

\bibitem[{{Peacock} \& {Smith}(2000)}]{Peacock2000}
{Peacock}, J.~A., \& {Smith}, R.~E. 2000, \mnras, 318, 1144,
  \dodoi{10.1046/j.1365-8711.2000.03779.x}

\bibitem[{{Planck Collaboration} {et~al.}(2020){Planck Collaboration},
  {Aghanim}, {Akrami}, {Ashdown}, {Aumont}, {Baccigalupi}, {Ballardini},
  {Banday}, {Barreiro}, {Bartolo}, {Basak}, {Battye}, {Benabed}, {Bernard},
  {Bersanelli}, {Bielewicz}, {Bock}, {Bond}, {Borrill}, {Bouchet}, {Boulanger},
  {Bucher}, {Burigana}, {Butler}, {Calabrese}, {Cardoso}, {Carron},
  {Challinor}, {Chiang}, {Chluba}, {Colombo}, {Combet}, {Contreras}, {Crill},
  {Cuttaia}, {de Bernardis}, {de Zotti}, {Delabrouille}, {Delouis}, {Di
  Valentino}, {Diego}, {Dor{\'e}}, {Douspis}, {Ducout}, {Dupac}, {Dusini},
  {Efstathiou}, {Elsner}, {En{\ss}lin}, {Eriksen}, {Fantaye}, {Farhang},
  {Fergusson}, {Fernandez-Cobos}, {Finelli}, {Forastieri}, {Frailis},
  {Fraisse}, {Franceschi}, {Frolov}, {Galeotta}, {Galli}, {Ganga},
  {G{\'e}nova-Santos}, {Gerbino}, {Ghosh}, {Gonz{\'a}lez-Nuevo}, {G{\'o}rski},
  {Gratton}, {Gruppuso}, {Gudmundsson}, {Hamann}, {Handley}, {Hansen},
  {Herranz}, {Hildebrandt}, {Hivon}, {Huang}, {Jaffe}, {Jones}, {Karakci},
  {Keih{\"a}nen}, {Keskitalo}, {Kiiveri}, {Kim}, {Kisner}, {Knox},
  {Krachmalnicoff}, {Kunz}, {Kurki-Suonio}, {Lagache}, {Lamarre}, {Lasenby},
  {Lattanzi}, {Lawrence}, {Le Jeune}, {Lemos}, {Lesgourgues}, {Levrier},
  {Lewis}, {Liguori}, {Lilje}, {Lilley}, {Lindholm}, {L{\'o}pez-Caniego},
  {Lubin}, {Ma}, {Mac{\'\i}as-P{\'e}rez}, {Maggio}, {Maino}, {Mandolesi},
  {Mangilli}, {Marcos-Caballero}, {Maris}, {Martin}, {Martinelli},
  {Mart{\'\i}nez-Gonz{\'a}lez}, {Matarrese}, {Mauri}, {McEwen}, {Meinhold},
  {Melchiorri}, {Mennella}, {Migliaccio}, {Millea}, {Mitra},
  {Miville-Desch{\^e}nes}, {Molinari}, {Montier}, {Morgante}, {Moss}, {Natoli},
  {N{\o}rgaard-Nielsen}, {Pagano}, {Paoletti}, {Partridge}, {Patanchon},
  {Peiris}, {Perrotta}, {Pettorino}, {Piacentini}, {Polastri}, {Polenta},
  {Puget}, {Rachen}, {Reinecke}, {Remazeilles}, {Renzi}, {Rocha}, {Rosset},
  {Roudier}, {Rubi{\~n}o-Mart{\'\i}n}, {Ruiz-Granados}, {Salvati}, {Sandri},
  {Savelainen}, {Scott}, {Shellard}, {Sirignano}, {Sirri}, {Spencer},
  {Sunyaev}, {Suur-Uski}, {Tauber}, {Tavagnacco}, {Tenti}, {Toffolatti},
  {Tomasi}, {Trombetti}, {Valenziano}, {Valiviita}, {Van Tent}, {Vibert},
  {Vielva}, {Villa}, {Vittorio}, {Wand elt}, {Wehus}, {White}, {White},
  {Zacchei}, \& {Zonca}}]{Planck2018}
{Planck Collaboration}, {Aghanim}, N., {Akrami}, Y., {et~al.} 2020, \aap, 641,
  A6, \dodoi{10.1051/0004-6361/201833910}

\bibitem[{{Raichoor} {et~al.}(2020){Raichoor}, {Eisenstein}, {Karim}, {Newman},
  {Moustakas}, {Brooks}, {Dawson}, {Dey}, {Duan}, {Eftekharzadeh},
  {Gazta{\~n}aga}, {Kehoe}, {Landriau}, {Lang}, {Lee}, {Levi}, {Meisner},
  {Myers}, {Palanque-Delabrouille}, {Poppett}, {Prada}, {Ross}, {Schlegel},
  {Schubnell}, {Staten}, {Tarl{\'e}}, {Tojeiro}, {Y{\`e}che}, \&
  {Zhou}}]{Raichoor2020}
{Raichoor}, A., {Eisenstein}, D.~J., {Karim}, T., {et~al.} 2020, Research Notes
  of the American Astronomical Society, 4, 180,
  \dodoi{10.3847/2515-5172/abc078}

\bibitem[{{Robotham} {et~al.}(2006){Robotham}, {Wallace}, {Phillipps}, \& {De
  Propris}}]{Robotham2006}
{Robotham}, A., {Wallace}, C., {Phillipps}, S., \& {De Propris}, R. 2006, \apj,
  652, 1077, \dodoi{10.1086/508130}

\bibitem[{{Rodriguez} \& {Merch{\'a}n}(2020)}]{Rodriguez2020}
{Rodriguez}, F., \& {Merch{\'a}n}, M. 2020, \aap, 636, A61,
  \dodoi{10.1051/0004-6361/201937423}

\bibitem[{Rodriguez {et~al.}(2015)Rodriguez, Merchán, \&
  Sgró}]{Rodriguez2015}
Rodriguez, F., Merchán, M., \& Sgró, M.~A. 2015, Astronomy \& Astrophysics,
  580, A86, \dodoi{10.1051/0004-6361/201525798}

\bibitem[{{Ruiz-Macias} {et~al.}(2020){Ruiz-Macias}, {Zarrouk}, {Cole},
  {Norberg}, {Baugh}, {Brooks}, {Dey}, {Duan}, {Eftekharzadeh}, {Eisenstein},
  {Forero-Romero}, {Gazta{\~n}aga}, {Hahn}, {Kehoe}, {Landriau}, {Lang},
  {Levi}, {Lucey}, {Meisner}, {Moustakas}, {Myers}, {Palanque-Delabrouille},
  {Poppett}, {Prada}, {Raichoor}, {Schlegel}, {Schubnell}, {Tarl{\'e}},
  {Weinberg}, {Wilson}, \& {Y{\`e}che}}]{Ruiz-Macias2020}
{Ruiz-Macias}, O., {Zarrouk}, P., {Cole}, S., {et~al.} 2020, Research Notes of
  the American Astronomical Society, 4, 187, \dodoi{10.3847/2515-5172/abc25a}

\bibitem[{{Rykoff} {et~al.}(2014){Rykoff}, {Rozo}, {Busha}, {Cunha},
  {Finoguenov}, {Evrard}, {Hao}, {Koester}, {Leauthaud}, {Nord}, {Pierre},
  {Reddick}, {Sadibekova}, {Sheldon}, \& {Wechsler}}]{Rykoff2014}
{Rykoff}, E.~S., {Rozo}, E., {Busha}, M.~T., {et~al.} 2014, \apj, 785, 104,
  \dodoi{10.1088/0004-637X/785/2/104}

\bibitem[{{Sheth} {et~al.}(2001){Sheth}, {Mo}, \& {Tormen}}]{Sheth2001}
{Sheth}, R.~K., {Mo}, H.~J., \& {Tormen}, G. 2001, \mnras, 323, 1,
  \dodoi{10.1046/j.1365-8711.2001.04006.x}

\bibitem[{{Springel}(2005)}]{Springel2005}
{Springel}, V. 2005, \mnras, 364, 1105,
  \dodoi{10.1111/j.1365-2966.2005.09655.x}

\bibitem[{{Szabo} {et~al.}(2011){Szabo}, {Pierpaoli}, {Dong}, {Pipino}, \&
  {Gunn}}]{Szabo2011}
{Szabo}, T., {Pierpaoli}, E., {Dong}, F., {Pipino}, A., \& {Gunn}, J. 2011,
  \apj, 736, 21, \dodoi{10.1088/0004-637X/736/1/21}

\bibitem[{{Tago} {et~al.}(2006){Tago}, {Einasto}, {Saar}, {Einasto},
  {Suhhonenko}, {J{\~o}eveer}, {Vennik}, {Hein{\"a}m{\"a}ki}, \&
  {Tucker}}]{Tago2006}
{Tago}, E., {Einasto}, J., {Saar}, E., {et~al.} 2006, Astronomische
  Nachrichten, 327, 365, \dodoi{10.1002/asna.200510536}

\bibitem[{{Tinker}(2020)}]{Tinker2020}
{Tinker}, J.~L. 2020, arXiv e-prints, arXiv:2010.02946.
\newblock \doarXiv{2010.02946}

\bibitem[{{Tinker} {et~al.}(2005){Tinker}, {Weinberg}, {Zheng}, \&
  {Zehavi}}]{Tinker2005}
{Tinker}, J.~L., {Weinberg}, D.~H., {Zheng}, Z., \& {Zehavi}, I. 2005, \apj,
  631, 41, \dodoi{10.1086/432084}

\bibitem[{{Tucker} {et~al.}(2000){Tucker}, {Oemler}, {Hashimoto}, {Shectman},
  {Kirshner}, {Lin}, {Landy}, {Schechter}, \& {Allam}}]{Tucker2000}
{Tucker}, D.~L., {Oemler}, Augustus, J., {Hashimoto}, Y., {et~al.} 2000, \apjs,
  130, 237, \dodoi{10.1086/317348}

\bibitem[{{van den Bosch} {et~al.}(2004){van den Bosch}, {Norberg}, {Mo}, \&
  {Yang}}]{vandenBosch2004}
{van den Bosch}, F.~C., {Norberg}, P., {Mo}, H.~J., \& {Yang}, X. 2004, \mnras,
  352, 1302, \dodoi{10.1111/j.1365-2966.2004.08021.x}

\bibitem[{{van den Bosch} {et~al.}(2005){van den Bosch}, {Weinmann}, {Yang},
  {Mo}, {Li}, \& {Jing}}]{vandenBosch2005}
{van den Bosch}, F.~C., {Weinmann}, S.~M., {Yang}, X., {et~al.} 2005, \mnras,
  361, 1203, \dodoi{10.1111/j.1365-2966.2005.09260.x}

\bibitem[{{van den Bosch} {et~al.}(2003){van den Bosch}, {Yang}, \&
  {Mo}}]{vandenBosch2003}
{van den Bosch}, F.~C., {Yang}, X., \& {Mo}, H.~J. 2003, \mnras, 340, 771,
  \dodoi{10.1046/j.1365-8711.2003.06335.x}

\bibitem[{{van den Bosch} {et~al.}(2007){van den Bosch}, {Yang}, {Mo},
  {Weinmann}, {Macci{\`o}}, {More}, {Cacciato}, {Skibba}, \&
  {Kang}}]{vandenBosch2007}
{van den Bosch}, F.~C., {Yang}, X., {Mo}, H.~J., {et~al.} 2007, \mnras, 376,
  841, \dodoi{10.1111/j.1365-2966.2007.11493.x}

\bibitem[{Vikram {et~al.}(2017)Vikram, Lidz, \& Jain}]{Vikram2017}
Vikram, V., Lidz, A., \& Jain, B. 2017, Monthly Notices of the Royal
  Astronomical Society, stw3311, \dodoi{10.1093/mnras/stw3311}

\bibitem[{Viola {et~al.}(2015)Viola, Cacciato, Brouwer, Kuijken, Hoekstra,
  Norberg, Robotham, van Uitert, Alpaslan, Baldry, Choi, de~Jong, Driver,
  Erben, Grado, Graham, Heymans, Hildebrandt, Hopkins, Irisarri, Joachimi,
  Loveday, Miller, Nakajima, Schneider, Sifón, \& Verdoes~Kleijn}]{Viola2015}
Viola, M., Cacciato, M., Brouwer, M., {et~al.} 2015, Monthly Notices of the
  Royal Astronomical Society, 452, 3529, \dodoi{10.1093/mnras/stv1447}

\bibitem[{{Wang} {et~al.}(2014){Wang}, {Mo}, {Yang}, {Jing}, \&
  {Lin}}]{Elucid2014}
{Wang}, H., {Mo}, H.~J., {Yang}, X., {Jing}, Y.~P., \& {Lin}, W.~P. 2014, \apj,
  794, 94, \dodoi{10.1088/0004-637X/794/1/94}

\bibitem[{Wang {et~al.}(2016)Wang, Mo, Yang, Zhang, Shi, Jing, Liu, Li, Kang,
  \& Gao}]{Wang2016}
Wang, H., Mo, H.~J., Yang, X., {et~al.} 2016, The Astrophysical Journal, 831,
  164, \dodoi{10.3847/0004-637X/831/2/164}

\bibitem[{Wang {et~al.}(2018)Wang, Mo, Chen, Yang, Yang, Wang, van~den Bosch,
  Jing, Kang, Lin, {et~al.}}]{Wang2018}
Wang, H., Mo, H., Chen, S., {et~al.} 2018, The Astrophysical Journal, 852, 31

\bibitem[{{Wang} {et~al.}(2018){Wang}, {Mo}, {Chen}, {Yang}, {Yang}, {Wang},
  {van den Bosch}, {Jing}, {Kang}, {Lin}, {Lim}, {Huang}, {Lu}, {Li}, {Cui},
  {Zhang}, {Tweed}, {Wei}, {Li}, \& {Shi}}]{Wang2018Elu}
{Wang}, H., {Mo}, H.~J., {Chen}, S., {et~al.} 2018, \apj, 852, 31,
  \dodoi{10.3847/1538-4357/aa9e01}

\bibitem[{Wang {et~al.}(2020)Wang, Mo, Li, Meng, \& Chen}]{Wang2020}
Wang, K., Mo, H.~J., Li, C., Meng, J., \& Chen, Y. 2020.
\newblock \doarXiv{2006.05426}

\bibitem[{{Wei} {et~al.}(2018){Wei}, {Li}, {Kang}, {Luo}, {Xia}, {Wang},
  {Yang}, {Wang}, {Jing}, {Mo}, {Lin}, {Wang}, {Li}, {Lu}, {Zhang}, {Lim},
  {Tweed}, \& {Cui}}]{Wei2018elu}
{Wei}, C., {Li}, G., {Kang}, X., {et~al.} 2018, \apj, 853, 25,
  \dodoi{10.3847/1538-4357/aaa40d}

\bibitem[{{Weinmann} {et~al.}(2006){Weinmann}, {van den Bosch}, {Yang}, \&
  {Mo}}]{Weinmann2006}
{Weinmann}, S.~M., {van den Bosch}, F.~C., {Yang}, X., \& {Mo}, H.~J. 2006,
  \mnras, 366, 2, \dodoi{10.1111/j.1365-2966.2005.09865.x}

\bibitem[{{Wen} {et~al.}(2012){Wen}, {Han}, \& {Liu}}]{Wen2012}
{Wen}, Z.~L., {Han}, J.~L., \& {Liu}, F.~S. 2012, \apjs, 199, 34,
  \dodoi{10.1088/0067-0049/199/2/34}

\bibitem[{{Willmer}(2018)}]{Willmer2018}
{Willmer}, C. N.~A. 2018, \apjs, 236, 47, \dodoi{10.3847/1538-4365/aabfdf}

\bibitem[{{Yang} {et~al.}(2005{\natexlab{a}}){Yang}, {Mo}, {Jing}, \& {van den
  Bosch}}]{Yang2005c}
{Yang}, X., {Mo}, H.~J., {Jing}, Y.~P., \& {van den Bosch}, F.~C.
  2005{\natexlab{a}}, \mnras, 358, 217,
  \dodoi{10.1111/j.1365-2966.2005.08801.x}

\bibitem[{{Yang} {et~al.}(2003){Yang}, {Mo}, \& {van den Bosch}}]{Yang2003}
{Yang}, X., {Mo}, H.~J., \& {van den Bosch}, F.~C. 2003, \mnras, 339, 1057,
  \dodoi{10.1046/j.1365-8711.2003.06254.x}

\bibitem[{{Yang} {et~al.}(2006){Yang}, {Mo}, \& {van den Bosch}}]{Yang2006}
---. 2006, \apjl, 638, L55, \dodoi{10.1086/501069}

\bibitem[{Yang {et~al.}(2008)Yang, Mo, \& van~den Bosch}]{Yang2008}
Yang, X., Mo, H.~J., \& van~den Bosch, F.~C. 2008, The Astrophysical Journal,
  676, 248, \dodoi{10.1086/528954}

\bibitem[{Yang {et~al.}(2009)Yang, Mo, \& van~den Bosch}]{Yang2009}
---. 2009, The Astrophysical Journal, 695, 900,
  \dodoi{10.1088/0004-637X/695/2/900}

\bibitem[{{Yang} {et~al.}(2005{\natexlab{b}}){Yang}, {Mo}, {van den Bosch}, \&
  {Jing}}]{Yang2005a}
{Yang}, X., {Mo}, H.~J., {van den Bosch}, F.~C., \& {Jing}, Y.~P.
  2005{\natexlab{b}}, \mnras, 356, 1293,
  \dodoi{10.1111/j.1365-2966.2005.08560.x}

\bibitem[{{Yang} {et~al.}(2005{\natexlab{c}}){Yang}, {Mo}, {van den Bosch}, \&
  {Jing}}]{Yang2005b}
---. 2005{\natexlab{c}}, \mnras, 357, 608,
  \dodoi{10.1111/j.1365-2966.2005.08667.x}

\bibitem[{{Yang} {et~al.}(2007){Yang}, {Mo}, {van den Bosch}, {Pasquali}, {Li},
  \& {Barden}}]{Yang2007}
{Yang}, X., {Mo}, H.~J., {van den Bosch}, F.~C., {et~al.} 2007, \apj, 671, 153,
  \dodoi{10.1086/522027}

\bibitem[{{Yang} {et~al.}(2005{\natexlab{d}}){Yang}, {Mo}, {van den Bosch},
  {Weinmann}, {Li}, \& {Jing}}]{Yang2005d}
---. 2005{\natexlab{d}}, \mnras, 362, 711,
  \dodoi{10.1111/j.1365-2966.2005.09351.x}

\bibitem[{{Yang} {et~al.}(2012){Yang}, {Mo}, {van den Bosch}, {Zhang}, \&
  {Han}}]{Yang2012}
{Yang}, X., {Mo}, H.~J., {van den Bosch}, F.~C., {Zhang}, Y., \& {Han}, J.
  2012, \apj, 752, 41, \dodoi{10.1088/0004-637X/752/1/41}

\bibitem[{{Yang} {et~al.}(2018){Yang}, {Zhang}, {Wang}, {Liu}, {Lu}, {Li},
  {Shi}, {Jing}, {Mo}, {van den Bosch}, {Kang}, {Cui}, {Guo}, {Li}, {Lim},
  {Lu}, {Luo}, {Wei}, \& {Yang}}]{Yang2018}
{Yang}, X., {Zhang}, Y., {Wang}, H., {et~al.} 2018, \apj, 860, 30,
  \dodoi{10.3847/1538-4357/aac2ce}

\bibitem[{{Y{\`e}che} {et~al.}(2020){Y{\`e}che}, {Palanque-Delabrouille},
  {Claveau}, {Brooks}, {Chaussidon}, {Davis}, {Dawson}, {Dey}, {Duan},
  {Eftekharzadeh}, {Eisenstein}, {Gazta{\~n}aga}, {Kehoe}, {Landriau}, {Lang},
  {Levi}, {Meisner}, {Myers}, {Newman}, {Poppett}, {Prada}, {Raichoor},
  {Schlegel}, {Schubnell}, {Staten}, {Tarl{\'e}}, \& {Zhou}}]{Yeche2020}
{Y{\`e}che}, C., {Palanque-Delabrouille}, N., {Claveau}, C.-A., {et~al.} 2020,
  Research Notes of the American Astronomical Society, 4, 179,
  \dodoi{10.3847/2515-5172/abc01a}

\bibitem[{{Zandivarez} {et~al.}(2006){Zandivarez}, {Mart{\'\i}nez}, \&
  {Merch{\'a}n}}]{Zandivarez2006}
{Zandivarez}, A., {Mart{\'\i}nez}, H.~J., \& {Merch{\'a}n}, M.~E. 2006, \apj,
  650, 137, \dodoi{10.1086/503894}

\bibitem[{{Zheng} {et~al.}(2005){Zheng}, {Berlind}, {Weinberg}, {Benson},
  {Baugh}, {Cole}, {Dav{\'e}}, {Frenk}, {Katz}, \& {Lacey}}]{Zheng2005}
{Zheng}, Z., {Berlind}, A.~A., {Weinberg}, D.~H., {et~al.} 2005, \apj, 633,
  791, \dodoi{10.1086/466510}

\bibitem[{{Zhou} {et~al.}(2020{\natexlab{a}}){Zhou}, {Newman}, {Mao},
  {Meisner}, {Moustakas}, {Myers}, {Prakash}, {Zentner}, {Brooks}, {Duan},
  {Landriau}, {Levi}, {Prada}, \& {Tarle}}]{Zhou2020}
{Zhou}, R., {Newman}, J.~A., {Mao}, Y.-Y., {et~al.} 2020{\natexlab{a}}, arXiv
  e-prints, arXiv:2001.06018.
\newblock \doarXiv{2001.06018}

\bibitem[{{Zhou} {et~al.}(2020{\natexlab{b}}){Zhou}, {Newman}, {Dawson},
  {Eisenstein}, {Brooks}, {Dey}, {Dey}, {Duan}, {Eftekharzadeh},
  {Gazta{\~n}aga}, {Kehoe}, {Landriau}, {Levi}, {Licquia}, {Meisner},
  {Moustakas}, {Myers}, {Palanque-Delabrouille}, {Poppett}, {Prada},
  {Raichoor}, {Schlegel}, {Schubnell}, {Staten}, {Tarl{\'e}}, \&
  {Y{\`e}che}}]{Zhou2020b}
{Zhou}, R., {Newman}, J.~A., {Dawson}, K.~S., {et~al.} 2020{\natexlab{b}},
  Research Notes of the American Astronomical Society, 4, 181,
  \dodoi{10.3847/2515-5172/abc0f4}

\bibitem[{{Zou} {et~al.}(2019){Zou}, {Gao}, {Zhou}, \& {Kong}}]{Zou2019}
{Zou}, H., {Gao}, J., {Zhou}, X., \& {Kong}, X. 2019, \apjs, 242, 8,
  \dodoi{10.3847/1538-4365/ab1847}

\end{thebibliography}

\label{lastpage}

\end{document}